\documentclass[12pt]{article}

\usepackage{amssymb,amsfonts,epsfig}

\def\ltwid{\mathrel{\raise.3ex\hbox{$<$\kern-.75em\lower1ex\hbox{$\sim$}}}}
\def \be{\begin{equation}}
\def \ee{\end{equation}}
\def \bea{\begin{eqnarray}}
\def \eea{\end{eqnarray}}

\def \a{\alpha}
\def \b{\beta}

\def \m{\mu}
\def \n{\nu}

\def \r{\rho}
\def \si{\sigma}

\def \ka{\kappa}
\def \d{\delta}
\def \D{\Delta}

\begin{document}

\begin{titlepage}

\begin{flushright}
UFIFT-QG-07-04
\end{flushright}

\vspace{2cm}

\begin{center}
{\bf Quantum Gravity Corrections to the One Loop Scalar
Self-Mass during Inflation}
\end{center}

\vspace{.5cm}

\begin{center}
E. O. Kahya$^{\dagger}$ and R. P. Woodard$^{\ddagger}$
\end{center}

\vspace{.5cm}

\begin{center}
\it{Department of Physics \\
University of Florida \\
Gainesville, FL 32611}
\end{center}

\vspace{1cm}

\begin{center}
ABSTRACT
\end{center}
We compute the one loop corrections from quantum gravity to the
self-mass-squared of a massless, minimally coupled scalar on a locally
de Sitter background. The calculation was done using dimensional
regularization and renormalized by subtracting fourth order BPHZ
counterterms. Our result should determine whether quantum gravitational
loop corrections can significantly alter the dynamics of a scalar inflaton.

\vspace{2cm}

\begin{flushleft}
\end{flushleft}

\vspace{.4cm}
\begin{flushleft}
$^{\dagger}$ e-mail: emre@phys.ufl.edu \\
$^{\ddagger}$ e-mail: woodard@phys.ufl.edu
\end{flushleft}
\end{titlepage}

\section{Introduction}

One can understand quantum loop effects as the reaction of classical 
field theory to virtual particles. Increasing the number density of
these virtual particles strengthens quantum effects. The expansion 
of spacetime tends to do this by trapping virtual pairs in the
Hubble flow and delaying their annihilation. During inflation the
effect is so strong that long wavelength massless virtual particles 
can persist forever. On the other hand, most massless particles 
possess classical conformal invariance, which causes the rate at
which they emerge from the vacuum to redshift so that the number 
density of virtual particles is not increased.

Massless, minimally coupled scalars and gravitons are unique in
possessing zero mass without classical conformal invariance. Inflation
results in a vast enhancement of quantum effects for these particles.
That is the origin of the primordial scalar \cite{Slava} and tensor 
\cite{Alexei} perturbations predicted by inflation \cite{MFB,LL}.
Weinberg has recently shown that loop corrections to these perturbations
are also enhanced, although not enough to make them observable
\cite{SW,Bua}.

Because the enhancement derives from long wavelength virtual particles,
the strongest effects come from nonderivative interactions. A massless,
minimally coupled scalar with a quartic self-interaction is pushed up
its potential by inflationary particle production, thereby inducing a
violation of the weak energy condition \cite{OW1,OW2} and a nonzero 
scalar mass \cite{BOW,KO}. The vacuum polarization from a charged, 
massless, minimally coupled scalar induces a nonzero photon mass
\cite{PTW,PW1} and a small negative shift in the vacuum energy 
\cite{PTsW1}. The inflationary creation of massless, minimally coupled 
scalars which are Yukawa-coupled to a massless fermion gives the fermion
mass \cite{PW2,GP} and induces a negative vacuum energy that grows 
without bound \cite{MW1}. And, more recently, a variety of other interesting
quantum loop effects due to scalar particles have been investigated
\cite{DS1,DS2,VMS,BH,Sloth}.

Gravitons have derivative interactions which weakens the enhancement
they experience. At one loop order quantum gravity gives only a constant
shift in the vacuum energy \cite{Ford,Fabio,TW1}. At two loops one
finds a secular reduction \cite{TW2} which might help explain why the
observed cosmological constant is so much smaller than the natural
scales of fundamental physics \cite{TW3}. (But see \cite{GT} for a 
different view \cite{TW4}.) The inflationary production of gravitons 
also induces a growing fermion field strength \cite{MW2}.

It is natural to wonder about the result of combining a massless,
minimally coupled scalar with gravity. If there are significant quantum
corrections they might have important consequences for inflation, 
although we stress that our scalar is a spectator to $\Lambda$-driven 
inflation. In this paper we shall compute its self-mass-squared at one
loop order. That is already a major task! In a subsequent work \cite{subs}
we will use the result to solve for the quantum-corrected scalar mode 
functions to see if the inflationary production of gravitons has a 
significant impact on scalar propagation.

In the next section we derive those Feynman rules we shall require.
The computation is done in section 3. In section 4 we first derive the
necessary BPHZ (Bogoliubov-Parasiuk-Hepp-Zimmerman) counterterms \cite{BPHZ}
and then use them to obtain a fully renormalized result. Because the 
effect we are seeking derives from infrared --- indeed, cosmological scale 
--- virtual particles, the ambiguity in the finite parts of these 
counterterms should not matter. It was possible to show this explicitly 
for the analogous impact of inflationary gravitons on massless fermions 
\cite{MW2}. Our conclusions comprise section 5.

\section{Feynman Rules}

To facilitate dimensional regularization we work in $D$ spacetime
dimensions. Our Lagrangian is,
\begin{equation}
\mathcal{L} \equiv -\frac12 \partial_{\mu} \phi \;
\partial_{\mu}\phi g^{\mu\nu} \sqrt{-g} + \frac1{16\pi G}
\Bigl( R - (D\!-\!2) \Lambda \Bigr) \sqrt{-g} \; . \label{Lag}
\end{equation}
Here $G$ is Newton's constant and $\Lambda \equiv (D-1) H^2$ is the 
cosmological constant. Because our scalar is a spectator to 
$\Lambda$-driven inflation, its background value is zero. Our 
background geometry is the conformal coordinate patch of 
$D$-dimensional de Sitter space,
\begin{equation}
ds^2 = a^2 \Bigl( -d\eta^2 + d\vec{x} \cdot d\vec{x} \Bigr) \qquad
{\rm where} \qquad a(\eta) = -\frac1{H \eta} \; .
\end{equation}

Perturbation theory is expressed using the graviton field $h_{\mu\nu}(x)$,
\begin{equation}
g_{\mu\nu}(x) \equiv a^2 \Bigl(\eta_{\mu\nu} + \kappa
h_{\mu\nu}(x)\Bigr) \qquad {\rm where} \qquad \kappa^2 \equiv 16
\pi G \; .
\end{equation}
The inverse metric and the volume element have the following expansions,
\begin{eqnarray}
{g}^{\mu\nu} & = & \frac1{a^2}\Bigl(\eta^{\mu\nu} - \kappa h^{\mu\nu} 
+ \kappa^2 h^{\mu}_{~\rho} h^{\rho\nu} - \dots \Bigr) \; , \\
\sqrt{-{g}} & = & a^D \Bigl(1 + \frac12 \kappa h + \frac18
\kappa^2 h^2 - \frac14 \kappa^2 h^{\rho\sigma} h_{\rho\sigma} +
\dots \Bigr) \; .
\end{eqnarray}
This computation requires the $\phi^2 h$ and $\phi^2 h^2$ interactions
which derive from expanding the scalar kinetic term,
\begin{eqnarray}
\lefteqn{-\frac12 \partial_{\mu} \phi \partial_{\nu} g^{\mu\nu} \sqrt{-g} 
= -\frac12 \partial_{\mu} \phi \partial_{\nu} \phi a^{D-2} 
\Biggl\{ \eta^{\m\n} -\ka h^{\m\n} + \frac12 \eta^{\m\n} \ka h 
+ \frac18 \eta^{\m\n}\ka^2 h^2 } \nonumber \\ 
& & \hspace{3.3cm} -\frac14 \eta^{\m\n} \ka^2 h^{\r\si} h_{\r\si} 
-\frac12 \ka^2 h h^{\m\n} + \ka^2 h^{\m\r} h_{\r}^{\n} + O(\kappa^3) 
\Biggr\} \; . \qquad
\end{eqnarray}

\begin{table}

\vbox{\tabskip=0pt \offinterlineskip
\def\tablerule{\noalign{\hrule}}
\halign to360pt {\strut#& \vrule#\tabskip=1em plus2em&
\hfil#\hfil& \vrule#& \hfil#\hfil& \vrule#\tabskip=0pt\cr
\tablerule \omit&height4pt&\omit&&\omit&\cr 
&& {\rm I} && $V_I^{\alpha\beta}$ & \cr 
\omit&height4pt&\omit&&\omit&\cr \tablerule
\omit&height2pt&\omit&&\omit&\cr && $1$ && $i\ka \, a^{D-2}
\, \partial^{\a}_{1}\partial^{\b}_{2}$ & \cr
\omit&height2pt&\omit&&\omit&\cr \tablerule
\omit&height2pt&\omit&&\omit&\cr && $2$ &&
$-\frac{i}2 \ka \, a^{D-2} \, \eta^{\a\b}
\partial_{1} \!\cdot\! \partial_{2}$ & \cr
\omit&height2pt&\omit&&\omit&\cr \tablerule}}

\caption{3-Point Vertex Operators $V_{I }^{ \alpha\beta}$ contracted into
$\phi_1 \phi_2 \, h_{\a\b}$.}

\label{3-Point}

\end{table}

We represent the 3-point and 4-point interaction terms as vertex
operators acting on the fields. For example, the first of the 3-point
vertices is,
\begin{equation}
-\frac12 \kappa a^{D-2} \partial_{\alpha} \phi \partial_{\beta} \phi 
h^{\alpha\beta} \qquad \Longrightarrow \qquad V_1^{\alpha\beta} = 
i \kappa a^{D-2} \partial_1^{\alpha} \partial_2^{\beta} \; .
\end{equation}
We number the fields ``1'', ``2'', ``3'', etc, starting with the 
two scalars and proceeding to the gravitons. Although we extract a 
factor of $\frac12$ for the two identical scalars, it is more efficient, 
for our computation, {\it not} to extract a similar factor of $\frac12$ 
for the identical gravitons of the 4-point vertices. Then we can 
dispense with the symmetry factor. So our first 4-point vertex is,
\begin{eqnarray}
-\frac{\ka^2}{16}\; a^{D-2}\partial^{\mu} \phi \; \partial_{\mu}
\phi \; h^2 \qquad \Longrightarrow \qquad U_1^{\alpha\beta\rho\sigma}
= - \frac{i}8 \ka^2 a^{D-2} \eta^{\a\b} \eta^{\r\si} \partial_{1} 
\!\cdot\! \partial_{2} \; .
\end{eqnarray}
The 3-point vertices are listed in Table~\ref{3-Point};
Table~\ref{4-Point} gives the 4-point vertices.

\begin{table}

\vbox{\tabskip=0pt \offinterlineskip
\def\tablerule{\noalign{\hrule}}
\halign to360pt {\strut#& \vrule#\tabskip=1em plus2em&
\hfil#\hfil& \vrule#& \hfil#\hfil& \vrule#\tabskip=0pt\cr
\tablerule \omit&height4pt&\omit&&\omit&\cr && {\rm I}
&& $U_I^{\alpha\beta\rho\sigma}$ & \cr 
\omit&height4pt&\omit&&\omit&\cr \tablerule
\omit&height2pt&\omit&&\omit&\cr && $1$ && $- \frac{i}8 \ka^2 
a^{D-2} \eta^{\a\b} \eta^{\r\si} \partial_{1} \!\cdot\!
\partial_{2}$ & \cr
\omit&height2pt&\omit&&\omit&\cr \tablerule
\omit&height2pt&\omit&&\omit&\cr && $2$ && $\frac{i}4 \ka^2
a^{D-2} \eta^{\a\r} \eta^{\b\si} \partial_{1} \!\cdot\!
\partial_{2}$ & \cr
\omit&height2pt&\omit&&\omit&\cr \tablerule
\omit&height2pt&\omit&&\omit&\cr && $3$ && $\frac{i}2
\ka^2 \, a^{D-2} \eta^{\a\b} \partial^{\r}_{1} \partial^{\si}_{2}$ & \cr
\omit&height2pt&\omit&&\omit&\cr \tablerule
\omit&height2pt&\omit&&\omit&\cr && $4$ && $-i
\ka^2 \, a^{D-2} \partial^{\a}_1 \eta^{\b\r} \partial^{\si}_2$ & \cr
\omit&height2pt&\omit&&\omit&\cr \tablerule}}

\caption{4-Point Vertex Operators $U_{I}^{ \alpha\beta\rho\sigma}$
contracted into $\phi_1 \phi_2 h_{\a\b} h_{\r\si}$.}

\label{4-Point}

\end{table}

Three notational conventions will simplify our discussion of 
propagators. The first is to denote the background geometry
with a hat,
\begin{equation}
\widehat{g}_{\mu\nu} = a^2 \eta_{\mu\nu} \quad , \quad 
\widehat{g}^{\mu\nu} = \frac1{a^2} \eta^{\mu\nu} \quad , \quad 
\sqrt{-\widehat{g}} = a^{D} \quad {\rm and} \quad \widehat{R} =
D (D\!-\!1) H^2 \; .
\end{equation}
Second, because time and space are treated differently in the
gauge we shall employ, it is useful to have expressions for the 
purely spatial parts of the Lorentz metric and the Kronecker delta,
\begin{equation}
\overline{\eta}_{\mu\nu} \equiv \eta_{\mu\nu} + \delta^0_{\mu}
\delta^0_{\nu} \qquad {\rm and} \qquad
\overline{\delta}^{\mu}_{\nu} \equiv \delta^{\mu}_{\nu} -
\delta_0^{\mu} \delta^0_{\nu} \; .
\end{equation}
Finally, the various propagators have simple expressions in terms 
of $y(x;x')$, a function of the de Sitter invariant length 
$\ell(x;x')$ from $x^{\mu}$ to $x^{\prime \mu}$,
\begin{equation}
y(x;x') = 4 \sin^2\Bigl( \frac12 H \ell(x;x')\Bigr) \; = a a' H^2
\Bigl\{ \Vert \vec{x} \!-\! \vec{x}' \Vert^2 - \Bigl(\vert \eta \!-\!
\eta'\vert \!-\! i \delta\Bigr)^2 \Bigr\} \; , \label{ydef}
\end{equation}
where $a \equiv a(\eta)$ and $a' \equiv a(\eta')$.

The massless minimally coupled scalar propagator obeys,
\be
\partial_{\mu} \Bigl(\sqrt{-\widehat{g}} \, \widehat{g}^{\mu\nu} 
\partial_{\nu}\Bigr) \; i  \D_A(x;x') \; = \; i\;\d^D (x-x'). 
\label{Aeqn}\ee
It has long been known that there is no de Sitter invariant 
solution \cite{AF}. The de-Sitter breaking solution which is 
relevant for cosmology is the one which preserves homogeneity and 
isotropy. This is known as the ``E(3)'' vacuum \cite{BA}, and the 
minimal solution takes the form \cite{OW1,OW2},
\begin{equation}
i\Delta_A(x;x') = A(y) + k \ln(a a') \qquad {\rm where} \qquad k
\equiv \frac{H^{D-2}}{(4\pi)^{\frac{D}2}} \frac{\Gamma(D\!-\!1)}{
\Gamma(\frac{D}2)} \; .
\end{equation}
The de Sitter invariant function $A(y)$ is \cite{OW2},
\begin{eqnarray}
\lefteqn{A(y) \equiv \frac{H^{D-2}}{(4\pi)^{\frac{D}2}} \Biggl\{
\frac{\Gamma(\frac{D}2 \!-\! 1)}{\frac{D}2 \!-\! 1}
\Bigl(\frac{4}{y}\Bigr)^{ \frac{D}2 -1} \!+\!
\frac{\Gamma(\frac{D}2 \!+\! 1)}{\frac{D}2 \!-\! 2}
\Bigl(\frac{4}{y} \Bigr)^{\frac{D}2-2} \!-\! \pi
\cot\Bigl(\frac{\pi D}2\Bigr)
\frac{\Gamma(D \!-\! 1)}{\Gamma(\frac{D}2)} } \nonumber \\
& & \hspace{.7cm} + \sum_{n=1}^{\infty} \Biggl[\frac1{n}
\frac{\Gamma(n \!+\! D \!-\! 1)}{\Gamma(n \!+\! \frac{D}2)}
\Bigl(\frac{y}4 \Bigr)^n \!\!\!\! - \frac1{n \!-\! \frac{D}2 \!+\!
2} \frac{\Gamma(n \!+\!  \frac{D}2 \!+\! 1)}{ \Gamma(n \!+\! 2)}
\Bigl(\frac{y}4 \Bigr)^{n - \frac{D}2 +2} \Biggr] \Biggr\} .
\qquad \label{A}
\end{eqnarray}

To get the graviton propagator, we add the following gauge fixing 
term to the invariant Lagrangian \cite{TW5},
\begin{equation}
\mathcal{L}_{\rm GF} = -\frac12 a^{D-2} \eta^{\mu\nu} F_{\mu} F_{\nu}
\; , \; F_{\mu} \equiv \eta^{\rho\sigma} \Bigl(h_{\mu\rho ,
\sigma} - \frac12 h_{\rho \sigma , \mu} + (D \!-\! 2) H a
h_{\mu \rho} \delta^0_{\sigma} \Bigr) . \label{GF}
\end{equation}
We can partially integrate the quadratic part of the gauge fixed
Lagrangian to put it in the form $\frac12 h^{\mu\nu} 
D_{\mu\nu}^{~~\rho \sigma} h_{\rho\sigma}$, where the kinetic
operator is,
\begin{eqnarray}
\lefteqn{D_{\mu\nu}^{~~\rho\sigma} \equiv \left\{ \frac12
\overline{\delta}_{ \mu}^{~(\rho}
\overline{\delta}_{\nu}^{~\sigma)} - \frac14 \eta_{\mu\nu}
\eta^{\rho\sigma} - \frac1{2(D\!-\!3)} \delta_{\mu}^0
\delta_{\nu}^0
\delta_0^{\rho} \delta_0^{\sigma} \right\} D_A } \nonumber \\
& & \hspace{3cm} + \delta^0_{(\mu}
\overline{\delta}_{\nu)}^{(\rho} \delta_0^{\sigma)} \, D_B +
\frac12 \Bigl(\frac{D\!-\!2}{D\!-\!3}\Bigr) \delta_{\mu}^0
\delta_{\nu}^0 \delta_0^{\rho} \delta_0^{\sigma} \, D_C \; ,
\qquad
\end{eqnarray}
The three scalar differential operators are,
\begin{eqnarray}
D_A & \equiv & \partial_{\mu} \Bigl(\sqrt{-\widehat{g}} \,
\widehat{g}^{\mu\nu} \partial_{\nu}\Bigr) \; , \\
D_B & \equiv & \partial_{\mu} \Bigl(\sqrt{-\widehat{g}} \,
\widehat{g}^{\mu\nu} \partial_{\nu}\Bigr) - \frac1{D} 
\Bigl(\frac{D\!-\!2}{D\!-\!1}\Bigr) \widehat{R} 
\sqrt{-\widehat{g}} \; , \\
D_C & \equiv & \partial_{\mu} \Bigl(\sqrt{-\widehat{g}} \,
\widehat{g}^{\mu\nu} \partial_{\nu}\Bigr) - \frac2{D}
\Bigl(\frac{D\!-\!3}{D\!-\!1}\Bigr) \widehat{R} 
\sqrt{-\widehat{g}} \; .
\end{eqnarray}

The graviton propagator in this gauge has the form of a sum of
constant tensor factors times scalar propagators,
\begin{equation}
i\Bigl[{}_{\mu\nu} \Delta_{\rho\sigma}\Bigr](x;x') =
\sum_{I=A,B,C} \Bigl[{}_{\mu\nu} T^I_{\rho\sigma}\Bigr]
i\Delta_I(x;x') \; . \label{gprop}
\end{equation}
We can get the scalar propagators by inverting the scalar kinetic
operators,
\begin{equation}
D_I \times i\Delta_I(x;x') = i \delta^D(x - x') \qquad {\rm for}
\qquad I = A,B,C \; . \label{sprops}
\end{equation}
The tensor factors are,
\begin{eqnarray}
\Bigl[{}_{\mu\nu} T^A_{\rho\sigma}\Bigr] & = & 2 \,
\overline{\eta}_{\mu (\rho} \overline{\eta}_{\sigma) \nu} -
\frac2{D\!-\! 3} \overline{\eta}_{\mu\nu}
\overline{\eta}_{\rho \sigma} \; , \label{T^A} \\
\Bigl[{}_{\mu\nu} T^B_{\rho\sigma}\Bigr] & = & -4 \delta^0_{(\mu}
\overline{\eta}_{\nu) (\rho} \delta^0_{\sigma)} \; , \label{T^B}\\
\Bigl[{}_{\mu\nu} T^C_{\rho\sigma}\Bigr] & = & \frac2{(D \!-\!2)
(D \!-\!3)} \Bigl[(D \!-\!3) \delta^0_{\mu} \delta^0_{\nu} +
\overline{\eta}_{\mu\nu}\Bigr] \Bigl[(D \!-\!3) \delta^0_{\rho}
\delta^0_{\sigma} + \overline{\eta}_{\rho \sigma}\Bigr] \;
\label{T^C}.
\end{eqnarray}
With these definitions and equation (\ref{sprops}) we can see that
the graviton propagator satisfies the following equation,
\begin{equation}
D_{\mu\nu}^{~~\rho\sigma} \times i\Bigl[{}_{\rho\sigma}
\Delta^{\alpha\beta} \Bigr](x;x') = \delta_{\mu}^{(\alpha}
\delta_{\nu}^{\beta)} i \delta^D(x-x') \; .
\end{equation}

The most singular part of the scalar propagator is the propagator for
a massless, conformally coupled scalar \cite{BD},
\begin{equation}
{i\Delta}_{\rm cf}(x;x') = \frac{H^{D-2}}{(4\pi)^{\frac{D}2}}
\Gamma\Bigl( \frac{D}2 \!-\! 1\Bigr)
\Bigl(\frac4{y}\Bigr)^{\frac{D}2-1} \; .
\end{equation}
The $A$-type propagator obeys the same equation as that of a
massless, minimally coupled scalar. The de Sitter invariant B-type
and $C$-type propagators are,
\begin{eqnarray}
\lefteqn{i \Delta_B(x;x') =  i \Delta_{\rm cf}(x;x') -
\frac{H^{D-2}}{(4 \pi)^{\frac{D}2}} \! \sum_{n=0}^{\infty}\!
\left\{\!  \frac{\Gamma(n \!+\! D \!-\! 2)}{\Gamma(n \!+\!
\frac{D}2)} \Bigl(\frac{y}4 \Bigr)^n \right. }
\nonumber \\
& & \hspace{6.5cm} \left. - \frac{\Gamma(n \!+\!
\frac{D}2)}{\Gamma(n \!+\! 2)} \Bigl( \frac{y}4 \Bigr)^{n -
\frac{D}2 +2} \!\right\} \! , \qquad
\label{DeltaB} \\
\lefteqn{i \Delta_C(x;x') =  i \Delta_{\rm cf}(x;x') +
\frac{H^{D-2}}{(4\pi)^{\frac{D}2}} \! \sum_{n=0}^{\infty}
\left\{\! (n\!+\!1) \frac{\Gamma(n \!+\! D \!-\! 3)}{\Gamma(n
\!+\! \frac{D}2)}
\Bigl(\frac{y}4 \Bigr)^n \right. } \nonumber \\
& & \hspace{4.5cm} \left. - \Bigl(n \!-\! \frac{D}2 \!+\!  3\Bigr)
\frac{ \Gamma(n \!+\! \frac{D}2 \!-\! 1)}{\Gamma(n \!+\! 2)}
\Bigl(\frac{y}4 \Bigr)^{n - \frac{D}2 +2} \!\right\} \! . \qquad
\label{DeltaC}
\end{eqnarray}
They can also be expressed as hypergeometric functions
\cite{CR,DC},
\begin{eqnarray}
i\Delta_B(x;x') & = & \frac{H^{D-2}}{(4\pi)^{\frac{D}2}}
\frac{\Gamma(D\!-\!2) \Gamma(1)}{\Gamma(\frac{D}2)} \,
\mbox{}_2F_1\Bigl(D\!-\!2,1;\frac{D}2;1 \!-\!
\frac{y}4\Bigr) \; , \label{FDB} \\
i\Delta_C(x;x') & = & \frac{H^{D-2}}{(4\pi)^{\frac{D}2}}
\frac{\Gamma(D\!-\!3) \Gamma(2)}{\Gamma(\frac{D}2)} \,
\mbox{}_2F_1\Bigl(D\!-\!3,2;\frac{D}2;1 \!-\! \frac{y}4\Bigr) \; .
\label{FDC}
\end{eqnarray}
These propagators might look complicated but they are actually
simple to use since the sums vanish in $D=4$, and every term in
these sums goes like a positive power of $y(x;x')$. Therefore only
a small number of terms in the sums can contribute when multiplied 
by a fixed divergence.

\section{One Loop Self-Mass-Squared}

This is the heart of the paper. We first evaluate the contribution
from the 4-point vertices of Table~\ref{4-Point}. Then we compute the 
vastly more difficult contributions from products of two 3-point vertices
from Table~\ref{3-Point}. We do not renormalize at this stage,
although we do take $D=4$ in finite terms. Renormalization is postponed 
until the next section.

\begin{figure}
\centerline{\epsfig{file=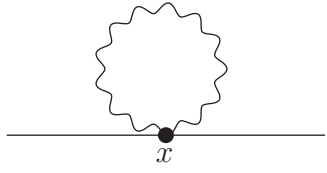}}
\caption{Contribution from 4-point vertices.}
\label{4pt}
\end{figure}

\subsection{Contributions from the 4-Point Vertices}

The generic diagram topology is depicted in Fig.~\ref{4pt}. The analytic
form is,
\begin{equation}
-i M^2_{\rm 4pt}(x;x') = \sum_{I=1}^4 U^{\alpha\beta \rho\sigma}_{I} 
\times i\Bigl[\mbox{}_{\alpha\beta} \Delta_{\rho\sigma} \Bigr](x;x) \,
\delta^D(x\!-\!x') \; . \label{4ptloop}
\end{equation}
In reading off the various contributions from Table~\ref{4-Point} one
should note that, whereas ``$\partial_2$'' acts upon $x^{\prime \mu}$,
the derivative operator ``$\partial_1$'' must be partially integrated
back onto the entire contribution. For example, the contribution from
$U_1^{\alpha\beta\rho\sigma}$ is,
\begin{eqnarray}
\lefteqn{-\frac{i}8 \kappa^2 a^{D-2} \eta^{\alpha\beta} \eta^{\rho\sigma}
\partial_1 \!\cdot\! \partial_2 \times i\Bigl[\mbox{}_{\alpha\beta} 
\Delta_{\rho\sigma} \Bigr](x;x) \, \delta^D(x\!-\!x') } \nonumber \\
& & \hspace{3cm} \Longrightarrow + \frac{i}8 \kappa^2 \partial^{\mu} 
\Biggl\{ a^{D-2} i\Bigl[\mbox{}^{\alpha}_{~ \alpha} \Delta^{\rho}_{~ 
\rho}\Bigr](x;x) \partial_{\mu}' \delta^D(x\!-\!x') \Biggr\} \; . \qquad
\end{eqnarray}
Reading off the other terms from Table~\ref{4-Point} gives,
\begin{eqnarray}
\lefteqn{-i M^2_{\rm 4pt}(x;x') = -\frac{i}8 \kappa^2 \partial^{\mu} 
\Biggl\{ a^{D-2} i\Bigl[\mbox{}^{\alpha}_{~ \alpha} \Delta^{\rho}_{~ 
\rho}\Bigr](x;x) \partial_{\mu} \delta^D(x\!-\!x') \Biggr\} 
+ \frac{i}4 \kappa^2 \partial^{\mu} \Biggl\{ a^{D-2} } \nonumber \\
& & \hspace{-.5cm} \times i\Bigl[\mbox{}^{\alpha\beta} \Delta_{\alpha\beta} 
\Bigr](x;x) \partial_{\mu} \delta^D(x\!-\!x') \Biggr\} + \frac{i}2
\kappa^2 \partial_{\rho} \Biggl\{ a^{D-2} i\Bigl[\mbox{}^{\alpha}_{~\alpha}
\Delta^{\rho\sigma} \Bigr](x;x) \partial_{\sigma} \delta^D(x\!-\!x') \Biggr\} 
\nonumber \\
& & \hspace{4.3cm} -i \kappa^2 \partial^{\alpha} \Biggl\{ a^{D-2} 
i\Bigl[\mbox{}_{\alpha\rho} \Delta^{\rho\sigma} \Bigr](x;x) \partial_{\sigma} 
\delta^D(x\!-\!x') \Biggr\} \; . \qquad \label{fourp}
\end{eqnarray}

It is apparent from expression (\ref{fourp}) that we require the 
coincidence limits of each of the three scalar propagators \cite{TW1},
\begin{eqnarray}
\lim_{x' \rightarrow x} i\Delta_A(x;x') & = & \frac{H^{D-2}}{
(4\pi)^{\frac{D}2}} \frac{\Gamma(D\!-\!1)}{\Gamma(\frac{D}2)} 
\Bigl\{-\pi \cot\Bigl(\frac{\pi}2 D\Bigr) + 2 \ln(a)\Bigr\} 
\; , \label{DAco} \qquad \\
\lim_{x' \rightarrow x} i\Delta_B(x;x') & = & \frac{H^{D-2}}{
(4\pi)^{\frac{D}2}} \frac{\Gamma(D\!-\!1)}{\Gamma(\frac{D}2)} 
\times -\frac1{D\!-\!2} \longrightarrow -\frac{H^2}{16 \pi^2} 
\; , \label{DBco} \\
\lim_{x' \rightarrow x} i\Delta_C(x;x') & = & \frac{H^{D-2}}{
(4\pi)^{\frac{D}2}} \frac{\Gamma(D\!-\!1)}{\Gamma(\frac{D}2)} 
\times \frac1{(D\!-\!2)(D\!-\!3)} \longrightarrow \frac{H^2}{16 \pi^2} 
\; . \label{DCco}
\end{eqnarray}
Note that the $B$-type and $C$-type propagators are finite for $D=4$. 
The four contractions of the coincident graviton propagator we require 
are \cite{TW1},
\begin{eqnarray}
i\Bigl[\mbox{}^{\alpha}_{~\alpha} \Delta^{\rho}_{\rho} \Bigr](x;x) 
& \longrightarrow & -4 \Bigl(\frac{D\!-\!1}{D\!-\!3}\Bigr) i\Delta_A(x;x)
+ 4 \times \frac{H^2}{16 \pi^2} \; , \qquad \\
i\Bigl[\mbox{}^{\alpha\beta} \Delta_{\alpha\beta} \Bigr](x;x) 
& \longrightarrow & \frac{(D\!-\!1)(D^2 \!-\! 3D \!-\! 2)}{D\!-\!3} \,
i\Delta_A(x;x) -2 \times \frac{H^2}{16 \pi^2} \; , \qquad \\
i\Bigl[\mbox{}^{\alpha}_{~\alpha} \Delta^{\rho\sigma} \Bigr](x;x) 
& \longrightarrow & - \frac4{D\!-\!3} \, \overline{\eta}^{\rho\sigma}
i\Delta_A(x;x) + \Bigl[2 \delta_0^{\rho} \delta_0^{\sigma} + 2
\overline{\eta}_{\rho\sigma} \Bigr] \times \frac{H^2}{16 \pi^2} 
\; , \qquad \\
i\Bigl[\mbox{}_{\alpha\rho} \Delta^{\rho\sigma} \Bigr](x;x) 
& \longrightarrow & \Bigl(\frac{D^2 \!-\! 3D \!-\! 2}{D\!-\!3}\Bigr) \,
\overline{\delta}^{\sigma}_{\alpha} i\Delta_A(x;x) + 2 \delta^0_{\alpha}
\delta^{\sigma}_0 \times \frac{H^2}{16 \pi^2} \; . \qquad
\end{eqnarray}
To save space we have taken $D=4$ in the finite contributions from
the $B$-type and $C$-type propagators.

Substituting these relations into expression (\ref{fourp}) and 
performing some trivial algebra gives the final result,
\begin{eqnarray}
\lefteqn{-i M^2_{\rm 4pt}(x;x') = \frac{i \kappa^2 H^{D-2}}{(4 \pi)^{
\frac{D}2}} \frac{\Gamma(D\!-\!1)}{\Gamma(\frac{D}2)} \times -\pi
\cot\Bigl(\frac{\pi}2 D\Bigr) \Biggl\{ \frac14 D (D\!-\!1) } \nonumber \\
& & \hspace{2.8cm} \times \partial^{\mu} \Bigl( a^{D-2} \partial_{\mu} 
\delta^D(x \!-\! x')\Bigr) - D a^{D-2} \nabla^2 \delta^D(x \!-\! x') 
\Biggr\} \nonumber \\
& & + \frac{i \kappa^2 H^2}{4 \pi^2} \Biggl\{ 3 \partial^{\mu}
\Bigl(a^2 \ln(a) \partial_{\mu} \delta^4(x \!-\! x')\Bigr)
- 4 \ln(a) a^2 \nabla^2 \delta^4(x \!-\! x') \nonumber \\
& & \hspace{2.8cm} - \partial^{\mu} \Bigl(a^2 \partial_{\mu} 
\delta^4(x \!-\!x') \Bigr) + a^2 \nabla^2 \delta^4(x \!-\! x')
\Biggr\} + O(D\!-\!4) \; . \qquad
\end{eqnarray}
Note that each of these terms vanishes in the flat space limit of
$H \rightarrow 0$ with the comoving time $t \equiv \ln(a)/H$ held fixed.
The reason for this is that the coincidence limit of the flat space
graviton propagator vanishes in dimensional regularization.

In order to combine $-iM^2_{\rm 4pt}$ with the 3-point contributions
it is useful to introduce notation for the scalar d'Alembertian in de 
Sitter background,
\begin{equation}
\square \equiv \frac1{\sqrt{-\widehat{g}}} \partial_{\mu} \Bigl(
\sqrt{-\widehat{g}} \, \widehat{g}^{\mu\nu} \partial_{\nu}\Bigr)
= \frac1{a^D} \partial^{\mu} \Bigl( a^{D-2} \partial_{\mu} \Bigr) \; .
\end{equation}
We also extract the logarithm from inside the d'Alembertian,
\begin{equation}
\partial^{\mu} \Bigl( a^2 \ln(a) \partial_{\mu} \delta^4(x \!-\!x') \Bigr)
= \frac12 \ln(a a') a^4 \square \delta^4(x \!-\!x') + \frac32 H^2 a^4
\delta^4(x \!-\! x') \; .
\end{equation}
With these conventions the final result takes the form,
\begin{eqnarray}
\lefteqn{-i M^2_{\rm 4pt}(x;x') = \frac{i \kappa^2 H^{D-2}}{(4 \pi)^{
\frac{D}2}} \frac{\Gamma(D\!-\!1)}{\Gamma(\frac{D}2)} \Biggl\{ \Bigl[
-\frac14 D (D\!-\!1) \pi \cot\Bigl(\frac{\pi}2 D\Bigr) - 2} \nonumber \\
& & \hspace{1cm} + 3 \ln(a a')\Bigr] a^D \square + \Bigl[ D \pi 
\cot\Bigl(\frac{\pi}2 D\Bigr) \!+\! 2 \!-\! 4 \ln(a a')\Bigr] a^{D-2} 
\nabla^2 \nonumber \\
& & \hspace{6cm} + 9 H^2 a^D + O(D\!-\! 4) \Biggr\} \delta^D(x \!-\! x') 
\; . \qquad \label{fin4pt}
\end{eqnarray}

\subsection{Contributions from the 3-Point Vertices}

\begin{figure}
\centerline{\epsfig{file=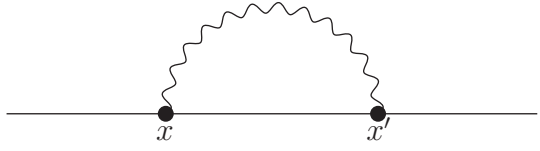}}
\caption{Contribution from two 3-point vertices.}
\label{3pt}
\end{figure}

In this section we calculate the contributions from two 3-point
vertex operators. It is diagrammatically represented in Fig.~\ref{3pt}.
Consulting Table~\ref{3-Point} and remembering to partially integrate
any derivative that acts upon an outer leg gives,
\begin{eqnarray}
\lefteqn{-iM^2_{\rm 3pt}(x;x') = \sum_{I=1}^2 V^{\alpha\beta }_{I}(x) 
\sum_{J=1}^2 V^{\rho\sigma }_{J}(x') \times i\Bigl[\mbox{}_{\alpha\beta} 
\Delta_{\rho\sigma}\Bigr](x;x') \, i\Delta_A(x;x') \; , } \\
& & = -\kappa^2 \partial_{\alpha} \partial_{\rho}' \Bigl\{ (a a')^{D-2}
i\Bigl[\mbox{}^{\alpha\beta} \Delta^{\rho\sigma}\Bigr]
\partial_{\beta} \partial_{\sigma}' i\Delta_A \Bigr\} + \frac{\kappa^2}2
\partial^{\mu} \partial_{\rho}' \Bigl\{ (a a')^{D-2} \nonumber \\
& & \hspace{1cm} \times i\Bigl[\mbox{}^{\alpha}_{~\alpha} \Delta^{\rho\sigma}
\Bigr] \partial_{\mu} \partial_{\sigma}' i\Delta_A \Bigr\} + \frac{\kappa^2}2 
\partial_{\alpha} \partial^{\prime \nu} \Bigl\{(a a')^{D-2} i\Bigl[\mbox{}^{
\alpha\beta} \Delta^{\rho}_{~\rho}\Bigr] \partial_{\beta} \partial_{\nu}' 
i\Delta_A \Bigr\} \nonumber \\
& & \hspace{5cm} - \frac{\kappa^2}4 \partial^{\mu} \partial^{\prime \nu} 
\Bigl\{ (a a')^{D-2} i\Bigl[\mbox{}^{\alpha}_{~\alpha} \Delta^{\rho}_{~\rho}
\Bigr] \partial_{\mu} \partial_{\nu}' i\Delta_A \Bigr\} \; . \qquad
\end{eqnarray}
Upon substituting the graviton propagator, performing the contractions
and segregating terms with the same scalar propagators, one finds three
generic sorts of terms. The first are those which involve two $A$-type
propagators,
\begin{eqnarray}
\lefteqn{\kappa^2 \nabla \!\cdot\! \nabla' \Bigl[ (a a')^{D-2} \;i\D_A \nabla
\!\cdot\! \nabla' \;i\D_A\Bigr] -\kappa^2 \Bigl(\frac{D\!-\!1}{D\!-\!3}\Bigr)
\partial_0 \partial_0' \Bigl[(a a')^{D-2} \;i\D_A\;
\partial_0 \partial_0' \;i\D_A\Bigr] } \nonumber \\
& & \hspace{0cm} + \kappa^2 \partial_i \partial_0' \Bigl[(a a')^{D-2} 
\;i\D_A\; \partial_i \partial_0' \;i\D_A\Bigr] + \kappa^2 \partial_0 
\partial_i' \Bigl[(a a')^{D-2}\;i\D_A\;
\partial_0 \partial_i' \;i\D_A\Bigr] \; . \qquad \label{AAs} \eea
The second kind of term involves one $A$-type and one $B$-type propagator,
\bea -\kappa^2 \partial_0 \partial_0'
\Bigl[(a a')^{D-2} \;i\D_B \nabla \!\cdot\! \nabla' \;i\D_A\Bigr]
- \kappa^2 \partial_i \partial_0' \Bigl[(a a')^{D-2} \;i\D_B\;
\partial_0 \partial_i' \;i\D_A\Bigr] \nonumber \\
- \kappa^2 \partial_0 \partial_i' \Bigl[(a a')^{D-2}\;i\D_B\;
\partial_i \partial_0' \;i\D_A\Bigr]
- \kappa^2 \nabla \!\cdot\! \nabla' \Bigl[ (a a')^{D-2} \;i\D_B\;
\partial_0 \partial_0' \;i\D_A\Bigr] \; . \label{BAs}
\eea 
Finally, there is the case of one propagator of $A$-type and the other
of $C$-type,
\be 2 \kappa^2 \Bigl(\frac{D\!-\!2}{D\!-\!3}\Bigr) \;
\partial_0 \partial_0' \Bigl[(a a')^{D-2} \;i\D_C \;
\partial_0 \partial_0' \;i\D_A\Bigr] \; . \label{CAs}
\ee

Each of the nine terms in expressions (\ref{AAs}-\ref{CAs}) has the form,
\begin{equation}
\kappa^2 \partial_{\mu} \partial_{\nu}' \Bigl[ (a a')^{D-2} i\Delta_{I}(x;x')
\partial_{\rho} \partial_{\sigma}' i\Delta_{A}(x;x')\Bigr] \; , \label{gen}
\end{equation}
where ``$I$'' might be $A$, $B$ or $C$. Note that the three propagators 
can be written almost entirely as functions of $y(x;x')$ defined in 
(\ref{ydef}),
\begin{equation}
i\Delta_{A}(x;x') = A(y) + k \ln(a a') \;\; , \;\;
i\Delta_{B}(x;x') = B(y) \;\; {\rm and} \;\; i\Delta_{C}(x;x') = C(y) \; .
\end{equation}
The functions $A(y)$, $B(y)$ and $C(y)$ can be read off from expressions
(\ref{A}), (\ref{DeltaB}) and (\ref{DeltaC}), respectively. Note also 
that the inner derivatives eliminate the de Sitter breaking term of
$i\Delta_A$,
\begin{equation}
\partial_{\rho} \partial_{\sigma}' i\Delta_A(x;x') = \delta^0_{\rho}
\delta^0_{\sigma} \frac{i}{a^{D-2}} \delta^D(x \!-\! x') + A''(y)
\frac{\partial y}{\partial x^{\rho}} \frac{\partial y}{\partial
x^{\prime \sigma}} + A'(y) \frac{\partial^2y}{\partial x^{\rho}
\partial x^{\prime \sigma}} \; . \label{d2A}
\end{equation}
It follows that the analysis breaks up into three parts:
\begin{itemize}
\item{{\it Local contributions} from the delta function in (\ref{d2A});}
\item{{\it Logarithm contributions} from the factor of $k \ln(a a')$ in 
the $A$-type propagator when $I = A$ in expression (\ref{gen}); and}
\item{{\it Normal contributions} to expression (\ref{gen}) of the form,
\begin{equation}
\kappa^2 \partial_{\mu} \partial_{\nu}' \Biggl\{ (a a')^{D-2} I(y)
\Biggl[A'' \frac{\partial y}{\partial x^{\rho}} \frac{\partial y}{\partial 
x^{\prime \sigma}} + A' \frac{\partial^2 y}{\partial x^{\rho} \partial 
x^{\prime \sigma}} \Biggr] \Biggr\} \; . \label{normal}
\end{equation}}
\end{itemize}
We shall devote a separate part of this subsection to each.

\subsubsection{Local Contributions}

These are the simplest contributions. They only come from the 2nd term
of (\ref{AAs}), the 4th term of (\ref{BAs}) and from (\ref{CAs}). To 
avoid overlap with the logarithm contributions of the next part we define 
the local contribution from the 4th term of (\ref{AAs}) without the
logarithm,
\begin{eqnarray}
\lefteqn{-\kappa^2 \Bigl(\frac{D\!-\!1}{D\!-\!3}\Bigr) \partial_0 
\partial_0' \Bigl[ (a a')^{D-2} A(y) \times \frac{i}{a^{D-2}}
\delta^D(x\!-\!x') \Bigr]  = \frac{i \kappa^2 H^{D-2}}{(4 \pi)^{\frac{D}2}} 
\frac{\Gamma(D)}{(D\!-\!3) \Gamma(\frac{D}2)} } \nonumber \\
& & \hspace{2cm} \times -\pi \cot\Bigl(\frac{D}2 \pi\Bigr) \Biggl\{
-a^D \square \delta^D(x\!-\!x') + a^{D-2} \nabla^2 \delta^D(x\!-\!x')
\Biggr\} . \qquad
\end{eqnarray}
Note that we have chosen to convert primed derivatives into unprimed, and
to absorb the temporal derivatives into a covariant d'Alembertian $\square$,
\begin{eqnarray}
-\partial_0 \Bigl( a^{D-2} \partial_0' \delta^D(x\!-\!x')\Bigr)
&\!\!\!\! =\!\!\!\! & -\partial^{\mu} \Bigl( a^{D-2} \partial_{\mu} 
\delta^D(x\!-\!x') \Bigr) \!+\! a^{D-2} \nabla^2 \delta^D(x\!-\!x') 
\; , \qquad \\
& \!\!\!\!\equiv\!\!\!\! & -a^D \square \delta^D(x\!-\!x') + a^{D-2} \nabla^2 
\delta^D(x\!-\!x') \; .
\end{eqnarray}
This will facilitate renormalization.

The other two local contributions are finite. The 4th term of (\ref{BAs})
gives,
\begin{eqnarray}
\lefteqn{-\kappa^2 \nabla \!\cdot\! \nabla' \Bigl[ (a a')^{D-2} B(y) 
\times \frac{i}{a^{D-2}} \delta^D(x \!-\! x')\Bigr] } \nonumber \\
& & \hspace{4.5cm} = -\frac{i \kappa^2 H^2}{16 \pi^2} \times a^2 
\nabla^2 \delta^4(x \!-\! x') + O(D\!-\!4) \; . \qquad
\end{eqnarray}
And (\ref{CAs}) gives,
\begin{eqnarray}
\lefteqn{2 \kappa^2 \Bigl(\frac{D\!-\!2}{D\!-\!3}\Bigr) 
\partial_0 \partial_0' \Bigl[(a a')^{D-2} C(y) \times 
\frac{i}{a^{D-2}} \delta^D(x \!-\! x')\Bigr] } \nonumber \\
& & \hspace{2.2cm} = \frac{i \kappa^2 H^2}{4 \pi^2} \Biggl\{
a^4 \square \delta^4(x\!-\!x') - a^2 \nabla^2 \delta^4(x \!-\! x') 
\Biggr\} + O(D\!-\!4) \; . \qquad
\end{eqnarray}
Summing the three local contributions gives,
\begin{eqnarray}
\lefteqn{-i M^2_{{\rm 3pt} \atop {\rm loc}}(x;x') = \frac{i \kappa^2 
H^{D-2}}{(4 \pi)^{\frac{D}2}} \frac{\Gamma(D\!-\!1)}{\Gamma(\frac{D}2)} 
\Biggl\{ \Bigl[\Bigl(\frac{D\!-\!1}{D \!-\! 3}\Bigr) \pi \cot\Bigl(\frac{\pi}2 
D\Bigr) + 2\Bigr] a^D \square } \nonumber \\
& & \hspace{1cm} + \Bigl[ -\Bigl(\frac{D\!-\!1}{D\!-\!3}\Bigr) \pi 
\cot\Bigl(\frac{\pi}2 D\Bigr) \!-\! \frac72 \Bigr] a^{D-2} \nabla^2 
+ O(D\!-\! 4) \Biggr\} \delta^D(x \!-\! x') \; . \qquad \label{fin3loc}
\end{eqnarray}

\subsubsection{Logarithm Contributions}

These all come from expression (\ref{AAs}). They can be simplified by
using the propagator equation (\ref{Aeqn}),
\begin{eqnarray}
\partial_0 \Bigl(a^{D-2} \partial_0 i\Delta_A(x;x')\Bigr) & = & -i
\delta^D(x \!-\! x') + a^{D-2} \nabla^2 A(y) \; , \\
\partial_0' \Bigl(a^{\prime D-2} \partial_0' i\Delta_A(x;x')\Bigr) 
& = & -i \delta^D(x \!-\! x') + a^{\prime D-2} \nabla^2 A(y) \; .
\end{eqnarray}
One can also take the limit $D=4$ because all the logarithm contributions
are finite. For example, the function $A(y)$ is,
\begin{equation}
A(y) = \frac{H^2}{16 \pi^2} \Biggl\{ \frac{4}{y} - 2 \, \ln\Bigl(\frac{y}4
\Bigr) - 1 + O(D\!-\!4) \Biggr\} \; . \label{D4A}
\end{equation}

The first term of (\ref{AAs}) gives,
\begin{eqnarray}
\lefteqn{\kappa^2 \nabla \!\cdot\! \nabla' \Bigl[ (a a')^{D-2} \times
k \ln(a a') \times \nabla \!\cdot\! \nabla' i\Delta_A(x;x')\Bigr]} \nonumber \\
& & \hspace{4.5cm} = \frac{\kappa^2 H^2}{8 \pi^2} \, \ln(a a') (a a')^2
\nabla^4 A(y) + O(D\!-\!4) \; . \qquad
\end{eqnarray}
The second term of (\ref{AAs}) has the most complicated reduction,
\begin{eqnarray}
\lefteqn{-\kappa^2 \Bigl(\frac{D\!-\!1}{D\!-\!3}\Bigr) \partial_0 \partial_0' 
\Bigl[ (a a')^{D-2} \times k \ln(a a') \times \partial_0 \partial_0' 
i\Delta_A(x;x')\Bigr]} \nonumber \\
& & = \frac{i 3 \kappa^2 H^2}{8 \pi^2} \, \ln(a a') \Bigl\{ -a^4 \square 
+ 2 a^2 \nabla^2\Bigr\} \delta^4(x \!-\! x') \nonumber \\
& & \hspace{2cm} - \frac{3 \kappa^2 H^2}{8 \pi^2} \, \ln(a a') (a a')^2 
\nabla^4 A(y) - \frac{i 9 \kappa^2 H^4}{8 \pi^2} \, a^4 \delta^4(x\!-\!x')
\nonumber \\
& & \hspace{3.5cm} - \frac{3 \kappa^2 H^3}{8 \pi^2} \, (a a')^2 (a \partial_0 
\!+\! a' \partial_0') \nabla^2 A(y) + O(D\!-\!4) \; . \qquad
\end{eqnarray}
The third term of (\ref{AAs}) gives,
\begin{eqnarray}
\lefteqn{\kappa^2 \partial_i \partial_0' \Bigl[ (a a')^{D-2} \times 
k \ln(a a') \times \partial_i \partial_0' i\Delta_A(x;x')\Bigr]} \nonumber \\
& & \hspace{1cm} = -\frac{i \kappa^2 H^2}{8 \pi^2} \, \ln(a a') a^2 
\nabla^2 \delta^4(x\!-\!x') + \frac{\kappa^2 H^2}{8 \pi^2} \, \ln(a a')
(a a')^2 \nabla^4 A(y) \nonumber \\
& & \hspace{5cm} + \frac{\kappa^2 H^3}{8 \pi^2} \, (a a')^2 a' \partial_0'
\nabla^2 A(y) + O(D\!-\!4) \; . \qquad
\end{eqnarray}
A very similar contribution derives from the final term of (\ref{AAs}),
\begin{eqnarray}
\lefteqn{\kappa^2 \partial_0 \partial_i' \Bigl[ (a a')^{D-2} \times 
k \ln(a a') \times \partial_0 \partial_i' i\Delta_A(x;x')\Bigr]} \nonumber \\
& & \hspace{1cm} = -\frac{i \kappa^2 H^2}{8 \pi^2} \, \ln(a a') a^2 
\nabla^2 \delta^4(x\!-\!x') + \frac{\kappa^2 H^2}{8 \pi^2} \, \ln(a a')
(a a')^2 \nabla^4 A(y) \nonumber \\
& & \hspace{5cm} + \frac{\kappa^2 H^3}{8 \pi^2} \, (a a')^2 a \partial_0
\nabla^2 A(y) + O(D\!-\!4) \; . \qquad
\end{eqnarray}

Combining all four terms results in some significant cancellations,
\begin{eqnarray}
\lefteqn{-i M^2_{{\rm 3pt} \atop {\rm log}}(x;x') = \frac{\kappa^2 H^2}{8 
\pi^2} \Biggl\{ \ln(a a') \Bigl[-3 a^4 \square + 4 a^2 \nabla^2\Bigr] i
\delta^4(x \!-\! x') } \nonumber \\
& & \hspace{.9cm} - 9 H^2 a^4 i \delta^4(x \!-\! x') - 2 H (a a')^2 
(a \partial_0 \!+\! a' \partial_0') \nabla^2 A(y) + O(D \!-\! 4) \Biggr\} .
\qquad \label{fin3log}
\end{eqnarray}
Each of the local terms in (\ref{fin3log}) cancels a similar finite, local 
4-point contribution in (\ref{fin4pt}), leaving only the nonlocal
contribution involving derivatives of $A(y)$. It is possible to eliminate 
the temporal derivatives in this expression. However, the procedure is best
explained in the final part of this subsection.

\subsubsection{Normal Contributions}

These contributions are the most challenging. Our strategy for reducing
them is to first extract the $\partial_{\rho}$ and $\partial_{\sigma}'$ 
derivatives from (\ref{normal}) {\it generically}, without exploiting the 
functional forms of $A(y)$, $B(y)$ and $C(y)$. We also convert all primed 
derivatives into unprimed ones and express the final result in terms of 
ten ``External Operators''. This not only makes it possible to perceive 
general relations, it also reduces the superficial degree of divergence 
of the terms we must eventually expand. And it leaves functions of the de 
Sitter invariant variable $y(x;x')$ for which an improved expansion 
procedure is possible \cite{PTsW2}.

This step of extracting derivatives is still quite involved so we shall
describe only the essentials in the body of the paper and consign the 
details to an appendix. The appendix also gives tabulated results for 
each of the ten External Operators. The final reduction of these generic 
tabulated results is straightforward. This subsection closes with a 
description of the technique and a pair of tables giving the final 
potentially divergent and manifestly finite contributions, respectively.

Our generic method for extracting derivatives requires one to carry out 
many indefinite integrations of functions of $y$. We define this
operation by the symbol $I[f](y)$,
\begin{equation}
I[f](y) \equiv \int^y dy' f(y') \; .
\end{equation}
If the function $F(y)$ is the product of two propagator functions, then
acting two derivatives on it can never produce a delta function,
\begin{equation}
\partial_{\rho} \partial_{\sigma}' F(y) = F''(y) \frac{\partial y}{
\partial x^{\rho}} \frac{\partial y}{\partial x^{\prime \sigma}} +
F'(y) \frac{\partial^2 y}{\partial x^{\rho} \partial x^{\prime \sigma}} \; . 
\end{equation}
It follows that we can express the inner part of the basic normal 
contribution (\ref{normal}) in terms of integrals of such products,
\begin{equation}
f(y) \Biggl\{A''(y) \frac{\partial y}{\partial x^{\rho}} \frac{\partial y}{
\partial x^{\prime \sigma}} \!+\! A'(y) \frac{\partial^2 y}{\partial x^{\rho}
\partial x^{\prime \sigma}} \Biggr\} \!=\!
\partial_{\rho} \partial_{\sigma}' I^2[f A''](y) \!+\! \frac{\partial^2 y}{
\partial x^{\rho} \partial x^{\prime \sigma}} I[f' A'](y) \; . \label{keyI}
\end{equation}

We must still deal with the final term of (\ref{keyI}). In conformal
coordinates the mixed second derivative of $y(x;x')$ is \cite{KW1},
\begin{equation}
\frac{\partial^2 y}{\partial x^{\rho} \partial x^{\prime \sigma}}
= H^2 a a' \Bigl\{ y \delta^0_{\rho} \delta^0_{\sigma} - 2 a
\delta^0_{\rho} H \Delta x_{\sigma} + 2 a' H \Delta x_{\rho}
\delta^0_{\sigma} - 2 \eta_{\rho\sigma}\Bigr\}.
\end{equation}
Breaking this up into spatial and temporal components gives,
\begin{eqnarray}
\frac{\partial^2 y}{\partial x^{0} \partial x^{\prime 0}} \!=\!
H^2 a a' \Bigl[2 \!-\! y \!+\! 2 a a' H^2 \Vert \Delta
\vec{x}\Vert^2\Bigr] &\!\!\! ,\!\!\! & \frac{\partial^2
y}{\partial x^{0} \partial x^{\prime j}} \!=\!
H^2 a a' \!\times\! -2 a H \Delta x_j \; , \qquad \\
\frac{\partial^2 y}{\partial x^{i} \partial x^{\prime 0}} \!=\!
H^2 a a' \times 2 a' H \Delta x_i &\!\!\! ,\!\!\! &
\frac{\partial^2 y}{\partial x^{i}
\partial x^{\prime j}} \!=\! H^2 a a' \!\times\! -2 \eta_{ij} \; . \qquad
\end{eqnarray}
One consequence is,
\begin{equation}
a a' H^2 \Vert \Delta \vec{x} \Vert^2 f(y) = -\frac12 (D\!-\!1)
I[f](y) - \frac{\nabla \cdot \nabla'}{4 a a' H^2}
I^2[f](y) \; .
\end{equation}
Another consequence is the relations,\footnote{On the left hand side of
relation (\ref{ID013}) we mean the naive second derivative, {\it without}
the delta function.}
\begin{eqnarray}
f(y) \partial_0 \partial_0' A(y) & = & \partial_0 \partial_0'
I^2[f A''](y) -\frac12 \nabla \!\cdot\! \nabla' I^3[f' A'](y) 
\nonumber \\
& & \hspace{.5cm} + H^2 a a' \Bigl\{(2 \!-\! y) I[f'A'](y) -
(D\!-\!1) I^2[f' A'](y)\Bigr\} \; ,\label{ID013} \qquad \\
f(y) \partial_0 \partial_j' A(y) & = & \partial_0 \partial_j'
I^2[f A''](y) + H a \partial_j' I^2[f' A'](y) \; , \qquad \\
f(y) \partial_i \partial_0' A(y) & = & \partial_i \partial_0'
I^2[f A''](y) + H a' \partial_i I^2[f' A'](y) \; , \qquad \\
f(y) \partial_i \partial_j' A(y) & = & \partial_i \partial_j'
I^2[f A''](y) - 2 H^2 a a' \eta_{ij} I[f' A'](y) \; .
\label{ID016}\qquad
\end{eqnarray}

Using these identities it is possible to extract the derivatives
from the first of the $A$-terms,
\begin{eqnarray}
\lefteqn{\nabla \!\cdot\! \nabla' \Bigl[(a a')^{D-2}
A(y)
\nabla \!\cdot\! \nabla' A(y) \Bigr] } \nonumber \\
& & = (a a')^{D-2} (\nabla \!\cdot\! \nabla' )^2 I^2[A
A''](y) - 2 (D\!-\!1) H^2 (a a')^{D-1} \nabla \!\cdot\! \nabla'
I[A^{\prime 2}](y) \; , \qquad \\
& & = (a a')^{D-2} \nabla^4 I^2[A A''](y) + 2 (D\!-\!1) H^2 (a
a')^{D-1} \nabla^2 I[A^{\prime 2}](y) \; . \qquad
\end{eqnarray}
Only the first term in the expansion of $I^2[A A''](y)$
contributes a divergence; we can set $D \!=\! 4$ in the higher
terms. Similarly, only the first two terms in the expansion of
$I[A^{\prime 2}](y)$ can diverge.

It is very simple to convert the primed spatial derivatives to 
unprimed ones,
\begin{equation}
\partial_i' f(y) = -\partial_i f(y) \; .
\end{equation}
We already used this relation in reducing the first of the
$A$-terms. For time derivatives it is useful to note,
\begin{eqnarray}
\frac{\partial y}{\partial x^0} & = & H a \Bigl(y - 2 a' H \Delta
\eta\Bigr)
= H a \Bigl(y - 2 + 2 \frac{a'}{a}\Bigr) \; , \\
\frac{\partial y}{\partial x^{\prime 0}} & = & H a' \Bigl(y + 2 a
H \Delta \eta\Bigr) = H a' \Bigl(y - 2 + 2 \frac{a}{a'}\Bigr) \; .
\end{eqnarray}
From this follow three important identities. The simple one is,
\begin{equation}
\Bigl( \partial_0 + \partial_0'\Bigr) f(y) = H (a \!+\! a') y
f'(y) \; . \label{ID1}
\end{equation}
Another result is,
\begin{eqnarray}
\Bigl(a' \partial_0 + a \partial_0'\Bigr) f(y) & = & 2 H a a'
\times a a'
H^2 \Vert \Delta \vec{x} \Vert^2 f'(y) \; , \qquad \\
& = & -(D\!-\!1) H a a' f(y) + \frac{\nabla^2}{2 H} I[f](y) \; .
\qquad \label{ID2}
\end{eqnarray}
The final identity results from combining (\ref{ID1}) and
(\ref{ID2}),
\begin{eqnarray}
\Bigl(a \partial_0 + a' \partial_0'\Bigr) f(y) & \!\!\!=\!\!\! & (a \!+\!
a') (\partial_0 \!+\! \partial_0') f(y) - (a' \partial_0 \!+\! a
\partial_0') f(y)
\; , \\
& \!\!\!=\!\!\! & H (a \!+\! a')^2 y f'(y) \!+\! (D\!-\!1) H a a' f(y) 
\!-\! \frac{\nabla^2}{2 H} I[f](y) \; . \qquad \label{ID3}
\end{eqnarray}

We can now reduce the nonlocal logarithm contribution from equation
(\ref{fin3log}). Applying (\ref{ID3}) gives,
\begin{eqnarray}
\lefteqn{\frac{\kappa^2 H^2}{8 \pi^2} \times -2 H (a a')^2 (a \partial_0 
\!+\! a' \partial_0') \nabla^2 A(y) = \frac{\kappa^2 H^2}{16 \pi^2} 
\Biggl\{-12 (a a')^3 H^2 \nabla^2 A } \nonumber \\
& & \hspace{4cm} - 4 (a a')^2 (a \!+\! a')^2 \nabla^2 (y A') + 2 (a a')^2
\nabla^4 I[A]\Biggr\} . \qquad
\end{eqnarray}
The derivative and the integral are straightforward using the $D=4$
expansion for $A(y)$ given in (\ref{D4A}). The final result is reported in 
Table~\ref{LogNew}. Of course we have neglected terms which eventually 
vanish such as $\nabla^4 y$.

\begin{table}

\vbox{\tabskip=0pt \offinterlineskip
\def\tablerule{\noalign{\hrule}}
\halign to390pt {\strut#& \vrule#\tabskip=1em plus2em&
\hfil#\hfil& \vrule#& \hfil#\hfil& \vrule#\tabskip=0pt\cr
\tablerule \omit&height4pt&\omit&&\omit&\cr &&$\!\!\!\!{\rm
External \; Operator}\!\!\!\!$ && ${\rm Coefficient\ of}\;
\frac{ \kappa^2 H^4}{(4\pi)^4}$ & \cr
\omit&height4pt&\omit&&\omit&\cr 
\tablerule
\omit&height2pt&\omit&&\omit&\cr 
&& $(aa')^3 H^2 \nabla^2$ && $-\frac{12}{x} + 24 \ln{x}$ & \cr
\omit&height2pt&\omit&&\omit&\cr 
\tablerule
\omit&height2pt&\omit&&\omit&\cr 
&& $(aa')^2 (a + a')^2 H^2 \nabla^2$ && $\frac{4}{x}$ & \cr
\omit&height2pt&\omit&&\omit&\cr 
\tablerule
\omit&height2pt&\omit&&\omit&\cr 
&& $(aa')^2\nabla^4$ && $8 \ln{x} - 16 x \ln{x}$ & \cr
\omit&height2pt&\omit&&\omit&\cr 
\tablerule}}

\caption{Nonlocal Logarithm Contributions from relation (\ref{fin3log})
with $x \equiv \frac{y}{4}$.}

\label{LogNew}

\end{table}

We eventually want to absorb all double time derivatives into
covariant d'Alembertian's,
\begin{equation}
\square = -\frac1{a^2} \partial_0^2 - \frac{(D\!-\!2) H}{a}
\partial_0 + \frac1{a^2} \nabla^2 \; .
\end{equation}
This is most effectively done with the internal factors of $(a a')^{D-2}$. 
For example, consider reducing one of the mixed $A$-terms,
\begin{eqnarray}
\lefteqn{\partial_0 \partial_i' \Bigl[ (a a')^{D-2} A(y)
\partial_0 \partial_i' A(y) \Bigr] } \nonumber \\
& & = \nabla^{\prime 2} \partial_0 \Bigl[ (a a')^{D-2} \partial_0
I^2[A A''](y) \Bigr] + H \nabla^{\prime 2} \partial_0 \Bigl[ a^{D-1}
a^{\prime D-2} I^2[A^{\prime 2}](y) \Bigr] \; , \qquad \\
& & = -a^D a^{\prime D-2} \nabla^2 \square I^2[A A''](y) + (a
a')^{D-2} \nabla^4 I^2[A A''](y) \nonumber \\
& & \hspace{.5cm} + H a^{D-1} a^{\prime D-2} \nabla^2 \partial_0
I^2[A^{ \prime 2}](y) + (D\!-\!1) H^2 a^D a^{\prime D-2} \nabla^2
I^2[A^{\prime 2}](y) \; . \qquad  \label{ID11}
\end{eqnarray}
Note also that we can convert a primed covariant d'Alembertian to
an unprimed one if it acts on a function of just $y(x;x')$,
\begin{equation}
\square f(y) = H^2 \Bigl[(4y \!-\! y^2) f''(y) + D (2 \!-\! y)
f'(y)\Bigr] = \square' f(y) \; .
\end{equation}
This is used in reducing the other mixed $A$-term,
\begin{eqnarray}
\lefteqn{\partial_i \partial_0' \Bigl[ (a a')^{D-2} A(y)
\partial_i
\partial_0' A(y) \Bigr] } \nonumber \\
& & = \nabla^2 \partial_0' \Bigl[ (a a')^{D-2} \partial_0' I^2[A
A''](y) \Bigr] + H \nabla^2 \partial_0' \Bigl[ a^{D-2} a^{\prime
D-1} I^2[A^{\prime 2}](y)
\Bigr] \; , \qquad \\
& & = -a^{D-2} a^{\prime D} \nabla^2 \square' I^2[A A''](y) + (a
a')^{D-2}
\nabla^4 I^2[A A''](y) \nonumber \\
& & \hspace{.5cm} + H a^{D-2} a^{\prime D-1} \nabla^2 \partial_0'
I^2[A^{ \prime 2}](y) + (D\!-\!1) H^2 a^{D-2} a^{\prime D}
\nabla^2
I^2[A^{\prime 2}](y) \; , \label{ID14}\qquad \\
& & = -a^{D-2} a^{\prime D} \nabla^2 \square I^2[A A''](y) + (a
a')^{D-2}
\nabla^4 I^2[A A''](y) \nonumber \\
& & \hspace{.5cm} - H a^{D-3} a^{\prime D} \nabla^2 \partial_0
I^2[A^{ \prime 2}](y) + \frac12 a^{D-3} a^{\prime D-1} \nabla^4
I^3[A^{\prime 2}](y) \; . \qquad
\end{eqnarray}

The previous point can be summarized by the relations,
\begin{eqnarray}
\partial_0 \Bigl[ (a a')^{D-2} \partial_0 f(y)\Bigr] & = & - a^D a^{\prime D-2}
\square f(y) + (a a')^{D-2} \nabla^2 f(y) \; , \qquad \\
\partial_0' \Bigl[ (a a')^{D-2} \partial_0' f(y)\Bigr] & = & - a^{D-2}
a^{\prime D} \square f(y) + (a a')^{D-2} \nabla^2 f(y) \; . \qquad
\end{eqnarray}
Another important point is that it is almost always best to write
any single factor of the mixed product $\partial_0 \partial_0'$ as
follows,
\begin{equation}
\partial_0 \partial_0' = \frac12 (\partial_0 + \partial_0')^2 - \frac12
\partial_0^2 - \frac12 \partial_0^{\prime 2} \; .
\end{equation}
So we find the ubiquitous reduction,
\begin{eqnarray}
\lefteqn{\partial_0 \partial_0' \Bigl[(a a')^{D-2} f(y)\Bigr] = (a
a')^{D-2} \partial_0 \partial_0' f(y) } \nonumber \\
& & \hspace{.5cm} + (D\!-\!2) H (a a')^{D-2} (a' \partial_0 \!+\!
a \partial_0') f(y) + (D\!-\!2)^2 H^2 (a a')^{D-1} f(y) \; , \qquad \\
& & = \frac12 (a a')^{D-2} (a^2 \!+\! a^{\prime 2}) \Bigl[\square
f(y) \!+\! H^2 y f'(y)\Bigr] - (a a')^{D-2} \nabla^2 f(y) \nonumber \\
& & \hspace{.5cm} + \frac14 (D\!-\!2) (a a')^{D-2} \nabla^2
I[f](y) + \frac12 (D\!-\!2) (D\!-\!3) H^2 (a a')^{D-1} f(y) \nonumber \\
& & \hspace{3cm} + \frac12 H^2 (a \!+\! a')^2 (a a')^{D-2}
\Bigl[(D\!-\!1) y f'(y) \!+\! y^2 f''(y)\Bigr] \; . \qquad
\end{eqnarray}
Another example is the two $B$-terms,
\begin{eqnarray}
\lefteqn{\partial_i \partial_0' \Bigl[ (a a')^{D-2} B(y) \partial_0
\partial_i' A(y)\Bigr] + \partial_0 \partial_i' \Bigl[ (a a')^{D-2} B(y)
\partial_i \partial_0' A(y)\Bigr] } \nonumber \\
& & \hspace{-.5cm} = \partial_i (\partial_0 \!+\! \partial_0')
\Bigl[ (a a')^{D-2} B(y) (\partial_0 \!+\! \partial_0')
\partial_i' A(y)\Bigr] \nonumber \\
& & - \partial_i \partial_0 \Bigl[ (a a')^{D-2} B(y) \partial_0
\partial_i' A(y)\Bigr] - \partial_i \partial_0' \Bigl[ (a
a')^{D-2} B(y) \partial_0' \partial_i' A(y)\Bigr] \; , \qquad \label{line1} \\
& & \hspace{-.5cm} = -(a^2 \!+\! a^{\prime 2}) (a a')^{D-2} \nabla^2 
\Bigl\{\square I^2[A'' B](y) + H^2 I[A'B \!+\! yA''B](y)\Bigr\} \nonumber \\
& & + (a a')^{D-2} \nabla^4 \Bigl\{ 2 I^2[A'' B](y) - \frac12
I^3[A' B'](y)\Bigr\} - H^2 (a \!+\! a')^2 (a a')^{D-2} \nabla^2 \nonumber \\
& & \hspace{.2cm} \times \Bigl\{ (D\!-\!2) I[A' B \!+\! y A''
B](y) \!+\!  y A'(y) B(y) + y^2 A''(y) B(y) \nonumber \\
& & \hspace{.5cm} - y I[A'B'](y) \Bigr\} + (D\!-\!1) H^2 (a^2
\!+\! a a' \!+\! a^{\prime 2}) (a a')^{D-2} \nabla^2 I^2[A' B'](y)
\; . \qquad \label{line2}
\end{eqnarray}

Extracting derivatives in this way from the various normal contributions
results in functions of $y$ which are acted upon by ten external operators,
\begin{eqnarray}
\alpha & \equiv & (a a')^D \square^2 \; , \\
\beta & \equiv & (a a')^{D-1} (a^2 + a^{\prime 2}) H^2 \square \; , \\
\gamma_1 & \equiv & (a a')^D H^4 \; , \\
\gamma_2 & \equiv & (a a')^{D-1} (a^2 + a^{\prime 2}) H^4 \; , \\
\gamma_3 & \equiv & (a a')^{D-1} (a + a')^2 H^4 = 2 \gamma_1 + \gamma_2 \; , \\
\delta & \equiv & (a a')^{D-2} (a^2 + a^{\prime 2}) \nabla^2 \square \; , \\
\epsilon_1 & \equiv & (a a')^{D-1} H^2 \nabla^2 \; , \\
\epsilon_2 & \equiv & (a a')^{D-2} (a^2 + a^{\prime 2}) H^2 \nabla^2 \; , \\
\epsilon_3 & \equiv & (a a')^{D-2} (a + a')^2 H^2 \nabla^2 = 2
\epsilon_1
+ \epsilon_2 \; , \\
\zeta & \equiv & (a a')^{D-2} \nabla^4 \; .
\end{eqnarray}
Tables~\ref{alpha}-\ref{zeta} of the Appendix give explicit results for
each of these ten operators. Note that in addition to the three 
propagator functions $A(y)$, $B(y)$ and $C(y)$, we also employ the
following less singular differences:
\begin{equation}
\Delta B \equiv B - A \qquad {\rm and} \qquad \Delta C \equiv 2
\Bigl(\frac{D-2}{D-3} \Bigr) (C - A) \; .
\end{equation}

The next step is substituting the explicit forms (\ref{A}), (\ref{DeltaB}),
(\ref{DeltaC}) for the propagator functions into the results of
Tables~\ref{alpha}-\ref{zeta} and expanding to the required order. To
understand what this is, note that we will be integrating the result with
respect to $x^{\prime \mu}$ against a smooth function (the zeroth order 
mode solution) with the derivatives of the ``External Operators'' acted 
{\it outside} the integrals. Because $y(x;x')$ vanishes like $(x-x')^2$ at 
coincidence, it is only necessary to retain the dimensional regularization 
for terms which would go like $1/y^2$ and higher for $D=4$.

Although these tables involve a bewildering variety of different integrals
and derivatives, careful examination of the results shows that they derive 
from just eight products of the propagator functions,
\begin{equation}
A^{\prime 2} \; , \; A A'' \; , \; A' B' \; , \; A'' B \; , \; A' \Delta B'
\; , \; A'' \Delta B \; , \; A' \Delta C' \; {\rm and} \; A'' \Delta C \; .
\end{equation}
The most singular products of $A^{\prime 2}$ and $A A''$ always appear 
either doubly integrated --- e.g., $I^2[A A'']$ in Table~\ref{alpha} --- 
or else integrated once and then multiplied by $y$ --- e.g., $-\frac12 y I[A^{
\prime 2}]$ in Table~\ref{beta}. Hence we need only retain the dimensional
regularization for the $1/y^D$ terms of these expansions,
\begin{eqnarray}
\lefteqn{A^{\prime 2} = \frac{\Gamma^2(\frac{D}2)}{16} \frac{H^{2D-4}}{
(4\pi)^D} \Biggl\{ \Bigl(\frac{4}{y}\Bigr)^D + 4 \Bigl(\frac{4}{y}\Bigr)^3
+ 4 \Bigl(\frac{4}{y}\Bigr)^2 + O\Bigl(\frac{D\!-\!4}{y^3}\Bigr)\Biggr\}\; ,}\\
\lefteqn{A A'' = \frac{\Gamma^2(\frac{D}2)}{16} \frac{H^{2D-4}}{(4\pi)^D} 
\Biggl\{ \frac{D}{D\!-\!2} \Bigl(\frac{4}{y}\Bigr)^D - 4 \Bigl(\frac{4}{y}
\Bigr)^3 \ln\Bigl(\frac{y}{4}\Bigr) } \nonumber \\
& & \hspace{4.5cm} - 4 \Bigl(\frac{4}{y}\Bigr)^2 \ln\Bigl(\frac{y}4\Bigr)
- 2 \Bigl(\frac{4}{y}\Bigr)^2 + O\Bigl(\frac{D\!-\!4}{y^3}\Bigr) \Biggr\} 
\; . \qquad
\end{eqnarray}
The product $A' B'$ can appear with only a single integration --- e.g.,
$D I[A' B']$ in Table~\ref{beta} --- or multiplied by a single factor of $y$
--- e.g., $D y A' B'$ in Table~\ref{gamma2}. We must therefore retain the
dimensional regularization for the $1/y^{D-1}$ term,
\begin{equation}
A' B' = \frac{\Gamma^2(\frac{D}2)}{16} \frac{H^{2D-4}}{(4\pi)^D} \Biggl\{ 
\Bigl(\frac{4}{y}\Bigr)^D + (D\!-\!2) \Bigl(\frac{4}{y}\Bigr)^{D-1} + 
O\Bigl(\frac{D\!-\!4}{y^2}\Bigr) \Biggr\} \; .
\end{equation}
However, the product $A'' B$ is always shielded by two or more powers of
$y$, so the expansion we require for it is,
\begin{equation}
A'' B = \frac{\Gamma^2(\frac{D}2)}{16} \frac{H^{2D-4}}{(4\pi)^D} \Biggl\{ 
\frac{D}{D\!-\!2} \Bigl(\frac{4}{y}\Bigr)^D + 2 \Bigl(\frac{4}{y}\Bigr)^3 + 
O\Bigl(\frac{D\!-\!4}{y^3}\Bigr) \Biggr\} \; .
\end{equation}
The products involving $\Delta B$ and $\Delta C$ are less singular,
\begin{eqnarray}
\lefteqn{A' \Delta B' = \frac{\Gamma^2(\frac{D}2)}{16} \frac{H^{2D-4}}{
(4\pi)^D} \Biggl\{ -2\Bigl(\frac{4}{y}\Bigr)^{D-1} - 4 \Bigl(\frac{4}{y}
\Bigr)^2 + O\Bigl(\frac{D\!-\!4}{y^2}\Bigr) \Biggr\} \; , } \\
\lefteqn{A'' \Delta B = \frac{\Gamma^2(\frac{D}2)}{16} \frac{H^{2D-4}}{
(4\pi)^D} \Biggl\{ 4\Bigl(\frac{4}{y}\Bigr)^3 \ln\Bigl(\frac{y}4\Bigr) +
2 \Bigl(\frac{4}{y} \Bigr)^3 } \nonumber \\
& & \hspace{4.5cm} + 4 \Bigl(\frac{4}{y}\Bigr)^2 \ln\Bigl(\frac{y}4\Bigr) +
2 \Bigl(\frac{4}{y}\Bigr)^2 + O\Bigl(\frac{D\!-\!4}{y^3}\Bigr) \Biggr\} 
\; , \qquad \\
\lefteqn{A' \Delta C' = \frac{\Gamma^2(\frac{D}2)}{16} \frac{H^{2D-4}}{
(4\pi)^D} \Biggl\{ -8\Bigl(\frac{4}{y}\Bigr)^{D-1} - 16 \Bigl(\frac{4}{y}
\Bigr)^2 + O\Bigl(\frac{D\!-\!4}{y^2}\Bigr) \Biggr\} \; , } \\
\lefteqn{A'' \Delta C = \frac{\Gamma^2(\frac{D}2)}{16} \frac{H^{2D-4}}{
(4\pi)^D} \Biggl\{ 16\Bigl(\frac{4}{y}\Bigr)^3 \ln\Bigl(\frac{y}4\Bigr) +
8 \Bigl(\frac{4}{y} \Bigr)^3 } \nonumber \\
& & \hspace{4.5cm} + 16 \Bigl(\frac{4}{y}\Bigr)^2 \ln\Bigl(\frac{y}4\Bigr) +
8 \Bigl(\frac{4}{y}\Bigr)^2 + O\Bigl(\frac{D\!-\!4}{y^3}\Bigr) \Biggr\} 
\; . \qquad
\end{eqnarray}

\begin{table}
\vbox{\tabskip=0pt \offinterlineskip
\def\tablerule{\noalign{\hrule}}
\halign to390pt {\strut#& \vrule#\tabskip=1em plus2em&
\hfil#\hfil& \vrule#& \hfil#\hfil& \vrule#& \hfil#\hfil&
\vrule#\tabskip=0pt\cr \tablerule
\omit&height4pt&\omit&&\omit&&\omit&\cr
\omit&height2pt&\omit&&\omit&&\omit&\cr &&\omit\hidewidth $\;{\rm
Ext.\ Op.}$ \hidewidth && $\!\!\!\!{\rm Coef.\ of}\;
\frac{\kappa^2 H^{2D-4}}{(4\pi)^D} \Gamma^2(\frac{D}2) (\frac4{y})^{D-1}
\!\!\!\!$ && $\!\!\!\!{\rm Coef.\ of}\; \frac{\kappa^2 H^{2D-4}}{(4\pi)^D}
\Gamma^2(\frac{D}2) (\frac4{y})^{D-2}\!\!\!\!$ & \cr 
\omit&height4pt&\omit&&\omit&&\omit&\cr
\tablerule \omit&height2pt&\omit&&\omit&&\omit&\cr 
&& $\a$ && $0$ && $\frac{D}{(D-1)(D-2)^2}$ & \cr
\omit&height2pt&\omit&&\omit&&\omit&\cr \tablerule
\omit&height2pt&\omit&&\omit&&\omit&\cr 
&& $\b$ && $-\frac{D}{4(D-1)}$ && $-\frac{D^3-3D^2-4D+8}{4(D-1)(D-2)}$ & \cr
\omit&height2pt&\omit&&\omit&&\omit&\cr \tablerule
\omit&height2pt&\omit&&\omit&&\omit&\cr 
&& $\gamma_1$ && $-\frac{D(D-2)}{4}$ && $-\frac{D^3-3D^2-4D+8}{4}$ & \cr
\omit&height2pt&\omit&&\omit&&\omit&\cr \tablerule
\omit&height2pt&\omit&&\omit&&\omit&\cr 
&& $\gamma_2$ && $\frac{D}{4}$ && $\frac{D^3-3D^2-4D+8}{4(D-1)}$ & \cr
\omit&height2pt&\omit&&\omit&&\omit&\cr \tablerule
\omit&height2pt&\omit&&\omit&&\omit&\cr 
&& $\gamma_3$ && $-\frac{D}{4}$ && $0$ & \cr
\omit&height2pt&\omit&&\omit&&\omit&\cr \tablerule
\omit&height2pt&\omit&&\omit&&\omit&\cr 
&& $\d$ && $0$ && $0$ & \cr 
\omit&height2pt&\omit&&\omit&&\omit&\cr \tablerule
\omit&height2pt&\omit&&\omit&&\omit&\cr 
&& $\epsilon_1$ && $0$ && $\frac{(D^2-6D+4)}{2(D-1)(D-2)}$ & \cr
\omit&height2pt&\omit&&\omit&&\omit&\cr \tablerule
\omit&height2pt&\omit&&\omit&&\omit&\cr 
&& $\epsilon_2$ && $0$ && $\frac{(1-2D)}{(D-1)(D-2)}$ & \cr
\omit&height2pt&\omit&&\omit&&\omit&\cr \tablerule
\omit&height2pt&\omit&&\omit&&\omit&\cr 
&& $\epsilon_3$ && $0$ && $0$ & \cr 
\omit&height2pt&\omit&&\omit&&\omit&\cr \tablerule
\omit&height2pt&\omit&&\omit&&\omit&\cr && 
$\zeta$ && $0$ && $0$ & \cr 
\omit&height2pt&\omit&&\omit&&\omit&\cr \tablerule}}

\caption{Divergent Normal Contributions.}

\label{DivTerms}

\end{table}

One next substititues these expansions into the totals of 
Tables~\ref{alpha}-\ref{zeta} and performs the necessary integrations,
differentiations, multiplications and summations. We must also multiply by
the overall factor of $\kappa^2$. For example, the result for ``External
Operator'' $\alpha$ is,
\begin{eqnarray}
\lefteqn{\kappa^2 \Bigl\{I^2[A A''] + I^2[A'' \Delta C]\Bigr\} } \nonumber \\
& & = \frac{\Gamma^2(\frac{D}2)}{16} \frac{\kappa^2 H^{2D-4}}{(4\pi)^D} 
I^2\Biggl[ \frac{D}{D\!-\!2} \Bigl(\frac{4}{y}\Bigr)^D + 12 \Bigl(\frac{4}{y} 
\Bigr)^3 \ln\Bigl(\frac{y}{4}\Bigr) + 8 \Bigl(\frac{4}{y}\Bigr)^3 \nonumber \\
& & \hspace{4.5cm} +12 \Bigl(\frac{4}{y}\Bigr)^2 \ln\Bigl(\frac{y}4\Bigr)
+6 \Bigl(\frac{4}{y}\Bigr)^2 + O\Bigl(\frac{D\!-\!4}{y^3}\Bigr) \Biggr] 
\; , \qquad \\
& & = \frac{\kappa^2 H^{2D-4}}{(4\pi)^D} \Gamma^2\Bigl(\frac{D}2\Bigr)
\Biggl\{ \frac{D}{(D\!-\!1) (D\!-\!2)^2} \Bigl(\frac{4}{y}\Bigr)^{D-2} + 
6 \Bigl(\frac{4}{y} \Bigr) \ln\Bigl(\frac{y}{4}\Bigr) + 13 \Bigl(\frac{4}{y}
\Bigr) \nonumber \\
& & \hspace{4.5cm} -6 \ln^2\Bigl(\frac{y}4\Bigr) -18 \ln\Bigl(\frac{y}4\Bigr)
+ O\Bigl(\frac{D\!-\!4}{y}\Bigr) \Biggr\} \; . \qquad
\end{eqnarray}
We have tabulated the results for each of the ten ``External Operators''.
Table~\ref{DivTerms} gives the quadratically and logarithmically divergent
terms; Table~\ref{Finite Terms} gives the terms which are manifestly finite.
In all cases the expressions were worked out by hand and then checked with
Mathematica \cite{Wolfram}.

\begin{table}

\vbox{\tabskip=0pt \offinterlineskip
\def\tablerule{\noalign{\hrule}}
\halign to390pt {\strut#& \vrule#\tabskip=1em plus2em&
\hfil#\hfil& \vrule#& \hfil#\hfil& \vrule#\tabskip=0pt\cr
\tablerule \omit&height4pt&\omit&&\omit&\cr &&$\!\!\!\!{\rm Ext.\ Op.}
\!\!\!\!$ && $\!\!\!\!{\rm Coefficient\ of}\; \frac{\kappa^2 H^4}{(4 \pi)^4}
\!\!\!\!$ & \cr
\omit&height4pt&\omit&&\omit&\cr 
\tablerule
\omit&height2pt&\omit&&\omit&\cr 
&& $\a$ && $\frac{6\ln{x}}{x} + \frac{13}{x} -6\ln^2{x} -18 \ln{x}$ & \cr
\omit&height2pt&\omit&&\omit&\cr 
\tablerule
\omit&height2pt&\omit&&\omit&\cr 
&& $\b$ && $\frac{5}{x}-18\ln{x}$ & \cr 
\omit&height2pt&\omit&&\omit&\cr
\tablerule 
\omit&height2pt&\omit&&\omit&\cr 
&& $\gamma_1$ && $\frac{30}{x}-108\ln{x} - 36$ & \cr
\omit&height2pt&\omit&&\omit&\cr 
\tablerule
\omit&height2pt&\omit&&\omit&\cr 
&& $\gamma_2$ && $-\frac{5}{x} - 18$ & \cr 
\omit&height2pt&\omit&&\omit&\cr
\tablerule 
\omit&height2pt&\omit&&\omit&\cr 
&& $\gamma_3$ && $-\frac{5}{x} -36$ & \cr 
\omit&height2pt&\omit&&\omit&\cr
\tablerule 
\omit&height2pt&\omit&&\omit&\cr 
&& $\d$ && $-\frac{4\ln{x}}{x} -\frac{49}{6x} + 4\ln^2{x} +10\ln{x} 
+ 12x \ln{x}$ & \cr
\omit&height2pt&\omit&&\omit&\cr 
\tablerule
\omit&height2pt&\omit&&\omit&\cr 
&& $\epsilon_1$ && $-\frac{26}{3x} + 60\ln{x} -120x \ln{x} +72x$ & \cr
\omit&height2pt&\omit&&\omit&\cr 
\tablerule
\omit&height2pt&\omit&&\omit&\cr 
&& $\epsilon_2$ && $\frac{13}{6x} -12 \ln{x} + 12x \ln{x}$ & \cr
\omit&height2pt&\omit&&\omit&\cr 
\tablerule
\omit&height2pt&\omit&&\omit&\cr 
&& $\epsilon_3$ && $-\frac{11}{6x} + 36x \ln{x} +12x$ & \cr
\omit&height2pt&\omit&&\omit&\cr 
\tablerule
\omit&height2pt&\omit&&\omit&\cr 
&& $\zeta$ && $\frac{10\ln{x}}{3} - 24x\ln{x} +24x^2\ln{x} -36x^2$ & \cr
\omit&height2pt&\omit&&\omit&\cr 
\tablerule}}

\caption{Finite Normal Contributions in terms of $x \equiv \frac{y}{4}$.}

\label{Finite Terms}

\end{table}

\section{Renormalization}

\begin{figure}
\centerline{\epsfig{file=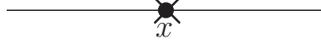}}
\caption{Contribution from counterterms.}
\label{ctms}
\end{figure}

In this section we obtain a completely finite result for the
self-mass-squared by subtracting 4th-order BPHZ counterterms \cite{BPHZ}.
We first identify two invariant counterterms which can contribute to 
this 1PI (One Particle Irreducible) function at one loop. Because our
gauge fixing functional (\ref{GF}) breaks de Sitter invariance \cite{TW5},
we must also consider noninvariant counterterms. We identify the only 
possible candidate based on a careful discussion of the residual symmetries
of our gauge fixing functional. It remains to collect and compute the
actual divergences. Contributions from the 4-point vertices are already
local, as are the ``local contributions'' from the 3-point vertices. Using a
now standard technique of partial integration \cite{OW1} we segregate the
divergences from the ``normal contributions'' of Table~\ref{DivTerms}.
In the end we identify the divergent parts of the three counterterms and
report a completely finite result.

One renormalizes the scalar self-mass-squared by subtracting diagrams 
of the form depicted in Fig.~\ref{ctms}. Because our scalar-graviton 
interactions have the form $\kappa^n h^n \partial \phi \partial \phi$, 
compared to the $\kappa^n h^n \partial h \partial h$ interactions of
pure gravity, the superficial degree of divergence at one loop order is
four, the same as that of pure quantum gravity. Of course the corresponding 
counterterms must contain two scalar fields, each of which has the 
dimension of a mass. Because we are dealing with one loop corrections from 
quantum gravity, all these counterterms must also carry a factor of the
loop counting parameter $\kappa^2 = 16 \pi G$, which has the dimension of 
an inverse mass-squared. Each counterterm must therefore have an additional 
mass dimension of four, either in the form of explicit masses or else as 
derivatives. The term with no derivatives is,
\begin{equation}
\kappa^2 m^4 \phi^2 \sqrt{-g} \; . \label{nods}
\end{equation}
There is no way to obtain an invariant with one derivative. Two derivatives
can act either on the scalars or on the metric to produce a curvature. We
can take the distinct terms to be,
\begin{equation}
\kappa^2 m^2 \partial_{\mu} \phi \partial_{\nu} \phi g^{\mu\nu} \sqrt{-g} 
\qquad {\rm and} \qquad \kappa^2 m^2 \phi^2 R \sqrt{-g} \; . \label{twods}
\end{equation}
There are no invariants with three derivatives. By judicious partial
integration and use of the Bianchi identity we can take the distinct terms
with four derivatives to be,
\begin{eqnarray}
\lefteqn{\kappa^2 \phi_{;\mu\nu} \phi_{;\rho\sigma} g^{\mu\nu} g^{\rho\sigma}
\sqrt{-g} \quad , \quad \kappa^2 \partial_{\mu} \phi \partial_{\nu} \phi R
g^{\mu\nu} \sqrt{-g} \quad , \quad \kappa^2 \partial_{\mu} \phi \partial_{\nu}
\phi R^{\mu\nu} \sqrt{-g} \; , } \nonumber \\
& & \hspace{3cm} \kappa^2 \phi^2 R^2 \sqrt{-g} \qquad {\rm and} \qquad
\kappa^2 \phi^2 R^{\mu\nu} R_{\mu\nu} \sqrt{-g} \; . \qquad \label{fourds}
\end{eqnarray}

Because our scalar is massless and mass is multiplicatively renormalized
in dimensional regularization, we can dispense with (\ref{nods}) and
(\ref{twods}). The last two counterterms of (\ref{fourds}) cannot occur
because the unrenormalized Lagragnian (\ref{Lag}) is invariant under $\phi 
\rightarrow \phi + {\rm const}$. The second and third terms of (\ref{fourds})
become degenerate when one uses the background equation, $\widehat{R}_{\mu\nu}
= (D\!-\!1) H^2 \widehat{g}_{\mu\nu}$. In the end just two independent
invariant counterterms survive, each with its own coefficient,
\begin{equation}
\frac12 \alpha_1 \kappa^2 \square \phi \square \phi a^D \qquad {\rm and}
\qquad -\frac12 \alpha_2 \kappa^2 H^2 \partial_{\mu} \phi \partial^{\mu} 
\phi a^{D-2} \; . \label{invar}
\end{equation}
The associated vertices are,
\begin{eqnarray}
\frac12 \alpha_1 \kappa^2 \square \phi \square \phi a^D & \longrightarrow &
i \alpha_1 \kappa^2 a^D \square^2 \delta^D(x\!-\!x') \; , \label{alp1} \\
-\frac12 \alpha_2 \kappa^2 H^2 \partial_{\mu} \phi \partial^{\mu} \phi a^{D-2}
& \longrightarrow & i \alpha_2 \kappa^2 H^2 a^D \square \delta^D(x \!-\! x')
\; . \label{alp2}
\end{eqnarray}

Had our gauge condition respected de Sitter invariance, all the divergences
in $-i M^2(x;x')$ could have been absorbed using (\ref{alp1}) and (\ref{alp2})
with appropriate choices for the divergent parts of the coefficients
$\alpha_1$ and $\alpha_2$. Although the reasons for it are not completely 
understood, there seems to be an obstacle to adding a de Sitter invariant 
gauge fixing functional \cite{AM,TW5,RPW}. This is why we employed the 
noninvariant functional (\ref{GF}). We must therefore describe how de Sitter 
transformations act in our conformal coordinate system and which
subgroup of them is respected by our gauge condition. The $\frac12 D
(D\!+\!1)$ de Sitter transformations can be decomposed as follows:
\begin{itemize}
\item{Spatial translations ---  $(D\!-\!1)$ distinct
transformations.}
\begin{eqnarray}
\eta' = \eta \; \;\;\;, \;\;\;\;  x^{\prime i} = x^i + \epsilon^i
\label{homog} \; .
\end{eqnarray}
\item{Rotations ---  $\frac12 (D\!-\!1) (D\!-\!2)$ distinct
transformations.}
\begin{eqnarray}
\eta' = \eta \;\;\;\; , \;\;\;\;x^{\prime i} = R^{ij} x^j
\label{isox} \; .
\end{eqnarray}
\item{Dilatation ---  $1$ distinct transformation.}
\begin{eqnarray}
\eta' = k \, \eta \;\;\;\; , \;\;\;\; x^{\prime i} = k \, x^i
\label{dilx} \; .
\end{eqnarray}
\item{Spatial special conformal transformations --- $(D\!-\!1)$
distinct transformations.}
\begin{eqnarray}
\eta' = \frac{\eta}{1 \!-\! 2 \vec{\theta} \!\cdot\! \vec{x} \!+\!
\Vert \vec{\theta} \Vert^2 x\!\cdot\! x} \;\;\; , \;\;\; x^{\prime
i} = \frac{x^i - \theta^i x\!\cdot\! x}{1 \!-\! 2 \vec{\theta}
\!\cdot\! \vec{x} \!+\! \Vert \vec{\theta} \Vert^2 x\!\cdot\! x}
\; . \label{sscx}
\end{eqnarray}
\end{itemize}

It turns out that our gauge choice breaks only spatial special
conformal transformations (\ref{sscx}) \cite{MW2}. Hence we can use the 
other symmetries to restrict possible noninvariant counterterms.
Spatial translational invariance means that there can be no dependence
upon $x^i$ except through the fields. Rotational invariance implies 
that spatial indices on derivatives must be contracted into one another.
Dilatation invariance implies that derivatives and the conformal time 
$\eta$ can only occur in the form $a^{-1} \partial_{\mu}$.

We can always use the invariant counterterms (\ref{alp1}-\ref{alp2}) to 
absorb a $\partial_0^2$ in favor of $\nabla^2$ and a single $\partial_0$,
\begin{equation}
\square = \frac1{a^2} \Bigl[ -\partial_0^2 - (D\!-\!2) H a
\partial_0 + \nabla^2\Bigr] \Longrightarrow \frac1{a^2}
\partial_0^2 = -\square -(D\!-\!2) \frac{H}{a} \partial_0 +
\frac{\nabla^2}{a^2} \; .
\end{equation}
We can also avoid $(\partial_0 \varphi)^2$,
\begin{equation}
\partial_{\mu} \varphi \partial_{\nu} \varphi g^{\mu\nu} =- \frac{(\partial_0
\varphi)^2}{a^2} + \frac1{a^2} \nabla \varphi \cdot
\nabla \varphi \Longrightarrow \frac{(\partial_0
\varphi)^2}{a^2} = -\partial_{\mu} \varphi
\partial_{\nu} \varphi g^{\mu\nu} + \frac1{a^2} \nabla \varphi
\cdot \nabla \varphi \; .
\end{equation}
One might think we need $H a^{D-1} \partial_0 \varphi \square
\varphi$, but a partial integration allows it to be written in
terms of an invariant counterterm and one with purely spatial
derivatives,
\begin{eqnarray}
\lefteqn{H a^{D-1} \partial_0 \varphi \square \varphi \longrightarrow 
-H a^{D-3} \partial_{\mu} \partial_0 \varphi \partial_{\nu}
\varphi \eta^{\mu\nu} - H^2 a^{D-2} (\partial_0 \varphi)^2 \; , } \\
& & \hspace{-.5cm} = -\frac12 H a^{D-3} \partial_0 (\partial_{\mu}
\varphi \partial_{\nu} \varphi ) \eta^{\mu\nu} + H^2 \partial_{\mu} \varphi
\partial_{\nu} \varphi g^{\mu\nu} \sqrt{-g} - H^2 a^{D-2} \nabla \varphi
\! \cdot \! \nabla \varphi \; , \qquad \\
& & \hspace{-.5cm} \longrightarrow \frac12 (D-1) H^2 \partial_{\mu} \varphi
\partial_{\nu} \varphi g^{\mu\nu} \sqrt{-g} - H^2 a^{D-2} \nabla
\varphi \cdot \nabla \varphi \; .
\end{eqnarray}
Another term one might consider is $H a^{D-3} \partial_0 \varphi
\nabla^2 \varphi$, but it can be partially integrated (twice) to give 
purely spatial deriviatives,
\begin{eqnarray}
H a^{D-3} \partial_0 \varphi \nabla^2 \varphi
& \longrightarrow & - H a^{D-3} \partial_0 \nabla \varphi
\cdot \nabla \varphi \; , \qquad \\
& = & -\frac12 H a^{D-3} \partial_0 (\nabla \varphi \cdot
\nabla \varphi) \; , \qquad \\
& \longrightarrow & \frac12 (D-3) H^2 a^{D-2} \nabla \varphi
\cdot \nabla \varphi \; . \qquad
\end{eqnarray}

Based on these considerations we conclude that only three noninvariant
counterterms might be needed in addition to the two invariant ones,
\begin{equation}
\frac12 \kappa^2 a^{D-2} \square \varphi \nabla^2 \varphi \;\; , \;\;
\frac12 \kappa^2 a^{D-4} \nabla^2 \varphi \nabla^2 \varphi \;\; {\rm and} 
\;\; -\frac12 \kappa^2 H^2 a^{D-2} \nabla \varphi \cdot \nabla \varphi \; . 
\label{noninv}
\end{equation}
Because our gauge fixing term (\ref{GF}) becomes Poincar\'e invariant in
the flat space limit of $H \rightarrow 0$ with the comoving time held fixed,
any noninvariant counterterm must vanish in this limit. Hence we require
only the final term of (\ref{noninv}). The vertex it gives is,
\begin{equation}
-\frac12 \alpha_3 \kappa^2 H^2 a^D \frac{\nabla}{a} \varphi \cdot
\frac{\nabla}{a} \varphi \longrightarrow i \alpha_3 \kappa^2 H^2
a^{D-2} \nabla^2 \delta^D(x\!-\!x') \; . \label{alp3}
\end{equation}

The structure of the three possible counterterms serves to guide our
further reduction of $-i M^2(x;x')$. First, we must convert all the factors
of $a'$ into $a$ on the local terms. Second, we see that factors of 
$H a^{D-3} \nabla^2 \partial_0 \delta^D(x-x')$ are not possible. Finally, 
it is not possible to get a divergence proportional to $H^3 a^{D-1} 
\partial_0 \delta^D(x\!-\!x')$ after using the delta function to convert 
all the factors of $a'$ into factors of $a$.

It is now time to collect the divergent terms from the previous two
sections. Those from the 4-point contributions, and from the ``local''
3-point contributions are already in a form which can be absorbed
into the three counterterms. However, we must still bring the ``normal''
3-point contributions of Table~\ref{DivTerms} to this form. Recall that
these terms involve powers of $y$ that are not integrable for $D = 4$
dimensions,
\begin{equation}
\Bigl(\frac{4}{y}\Bigr)^{D-1} \qquad {\rm and} \qquad
\Bigl(\frac{4}{y}\Bigr)^{D-2} \; .
\end{equation}
Our procedure is to extract d'Alembertians from these terms until they
become integrable using the identity,
\begin{eqnarray}
\lefteqn{\square f(y) = H^2 \Bigl[(4y \!-\! y^2) f''(y) + D (2 \!-\! y)
f'(y)\Bigr] } \nonumber \\
& & \hspace{4cm} + {\rm Res}\Bigl[y^{\frac{D}2-2} f\Bigr] \times
\frac{4 \pi^{\frac{D}2} H^{2-D}}{\Gamma(\frac{D}2 \!-\! 1)}
\frac{i}{\sqrt{-g}} \delta^D(x \!-\! x') \; . \qquad \label{key}
\end{eqnarray}
Here ${\rm Res}[F]$ stands for the residue of $F(y)$; that is, the
coefficient of $1/y$ in the Laurent expansion of the function
$F(y)$ around $y = 0$.

The key identity (\ref{key}) allows us to extract a covariant
d`Alembertian from each of the nonintegrable terms,
\begin{eqnarray}
\Bigl(\frac{4}{y}\Bigr)^{D-1} & = & \frac2{(D\!-\!2)^2}
\frac{\square}{H^2} \Bigl(\frac{4}{y}\Bigr)^{D-2} -
\frac2{D\!-\!2} \Bigl(\frac{4}{y}\Bigr)^{D-2}
\; , \qquad \label{1st} \\
\Bigl(\frac{4}{y}\Bigr)^{D-2} & = & \frac2{(D\!-\!3)(D\!-\!4)}
\frac{\square}{ H^2} \Bigl(\frac{4}{y}\Bigr)^{D-3} -
\frac4{D\!-\!4} \Bigl(\frac{4}{y}\Bigr)^{ D-3} \; . \label{2nd}
\qquad
\end{eqnarray}
We could use (\ref{2nd}) on (\ref{1st}) to reduce them both to the
power $1/y^{D-3}$. The power $1/y^{D-3}$ is integrable, so we could 
take $D \!=\! 4$ at this point were it not for the explicit factors of
$1/(D\!-\!4)$. 

To segregate the divergence on the local term we add zero in the form,
\begin{equation}
0 = \frac{\square}{H^2} \Bigl(\frac{4}{y}\Bigr)^{\frac{D}2-1} -
\frac{D}2 \Bigl(\frac{D}2 \!-\!1\Bigr)
\Bigl(\frac{4}{y}\Bigr)^{\frac{D}2-1} - \frac{(4\pi)^{\frac{D}2}
H^{-D}}{\Gamma(\frac{D}2 \!-\! 1)} \frac{i}{a^D} \delta^D(x
\!-\! x') \; . \label{delt}
\end{equation}
Using (\ref{delt}) in (\ref{2nd}) gives,
\begin{eqnarray}
\lefteqn{\Bigl(\frac{4}{y}\Bigr)^{D-2} \!\!\!= \frac{2}{(D \!-\!
3) (D\!-\!4)} \Biggl\{ \! \frac{(4\pi)^{\frac{D}2}
H^{-D}}{\Gamma(\frac{D}2 \!-\! 1)} \frac{i \delta^D(x \!-\!
x')}{a^D} \!+\!  \frac{\square}{H^2} \! \Biggl[ \!
\Bigl(\frac{4}{y} \Bigr)^{D-3} \!\!\! -
\Bigl(\frac{4}{y}\Bigr)^{\frac{D}2-
1} \! \Biggr] \! \Biggr\} } \nonumber \\
& & \hspace{4.5cm} - \frac{4}{D\!-\!4}
\Biggl\{\Bigl(\frac{4}{y}\Bigr)^{D-3} \!\!\! - \frac{D (D\!-\!
2)}{8 (D\!-\!3)} \Bigl(\frac{4}{y}\Bigr)^{\frac{D}2-1}
\Biggr\} \; , \qquad \\
& & = \frac{i H^{-D} (4\pi)^{\frac{D}2}}{(D\!-\!3)(D\!-\!4) 
\Gamma(\frac{D}2)} \times (D\!-\!2) \frac{\delta^D(x\!-\! x')}{a^D} 
- \frac{\square}{H^2} \Biggl\{ \frac{4}{y} \ln\Bigl(\frac{y}4\Bigr)
\Biggr\} \nonumber \\
& & \hspace{5.5cm} + 2 \Bigl(\frac{4}{y}\Bigr)
\ln\Bigl(\frac{y}{4}\Bigr) - \Bigl(\frac{4}{y} \Bigr) + O(D \!-\!
4) \; . \qquad \label{D-2}
\end{eqnarray}
The analogous result for the quadratically divergent term is,
\begin{eqnarray}
\lefteqn{\Bigl(\frac{4}{y}\Bigr)^{D-1} \!\!\!\!\! = \frac{i H^{-D} 
(4 \pi)^{\frac{D}2}}{(D \!-\! 3) (D\!-\!4) \Gamma(\frac{D}2)} 
\Biggl\{ \frac2{D\!-\!2} \frac{\square}{H^2} \!-\! 2\Biggr\}
\frac{\delta^D(x \!-\! x')}{a^D} - \frac12 \frac{\square^2}{H^4} 
\Biggl\{ \frac{4}{y} \ln\Bigl(\frac{y}4\Bigr) \Biggr\} } \nonumber \\
& & \hspace{.7cm} + \frac{\square}{H^2} \Biggl\{ 2
\Bigl(\frac{4}{y}\Bigr) \ln\Bigl(\frac{y}{4}\Bigr) \!-\! \frac12
\Bigl(\frac{4}{y} \Bigr)\Biggr\} \!-\! 2 \Bigl(\frac{4}{y}\Bigr)
\ln\Bigl(\frac{y}{4}\Bigr) \!+\! \Bigl(\frac{4}{y} \Bigr) \!+\!
O(D \!-\! 4) \; . \qquad \label{D-1}
\end{eqnarray}
The divergent local terms that result from applying (\ref{D-2}) and
(\ref{D-1}) to Table~\ref{DivTerms} are reported in Table~\ref{DivNew}.
Table~\ref{DivFin} gives the corresponding finite terms. In each case
we have eliminated the redundant External Operators $\gamma_3 = 2 \gamma_1 +
\gamma_2$ and $\epsilon_3 = 2 \epsilon_1 + \epsilon_2$.

\begin{table}

\vbox{\tabskip=0pt \offinterlineskip
\def\tablerule{\noalign{\hrule}}
\halign to390pt {\strut#& \vrule#\tabskip=1em plus2em&
\hfil#\hfil& \vrule#& \hfil#\hfil& \vrule#\tabskip=0pt\cr
\tablerule \omit&height4pt&\omit&&\omit&\cr &&$\!\!\!\!{\rm
External \; Operator}\!\!\!\!$ && ${\rm Coef.\ of}\;\;
\frac{i \kappa^2 H^{D-4}}{(4\pi)^{D/2}} \frac{\Gamma(\frac{D}{2})}{
(D-3)(D-4)} \frac{\delta^D(x- x')}{a^D}$ & \cr
\omit&height4pt&\omit&&\omit&\cr 
\tablerule
\omit&height2pt&\omit&&\omit&\cr 
&& $\alpha$ && $\frac{D}{(D-1)(D-2)}$ & \cr
\omit&height2pt&\omit&&\omit&\cr 
\tablerule
\omit&height2pt&\omit&&\omit&\cr 
&& $\beta$ && $-\frac{D}{2(D-1)(D-2)} \frac{\square}{H^2}
-\frac{(D+2)(D-4)}{4}$ & \cr
\omit&height2pt&\omit&&\omit&\cr 
\tablerule
\omit&height2pt&\omit&&\omit&\cr 
&& $\gamma_1$ && $-\frac{D^2}{2 (D-2)} \frac{\square}{H^2}
- \frac{(D^4-5D^3+16D-16)}{4}$ & \cr
\omit&height2pt&\omit&&\omit&\cr 
\tablerule
\omit&height2pt&\omit&&\omit&\cr 
&& $\gamma_2$ && $\frac{(D-2)(D^3-3D^2-4D+8)}{4 (D-1)}$ & \cr
\omit&height2pt&\omit&&\omit&\cr 
\tablerule
\omit&height2pt&\omit&&\omit&\cr 
&& $\delta$ && $0$ & \cr
\omit&height2pt&\omit&&\omit&\cr 
\tablerule
\omit&height2pt&\omit&&\omit&\cr 
&& $\epsilon_1$ && $\frac{(D^2-6D+4)}{2 (D-1)}$ & \cr
\omit&height2pt&\omit&&\omit&\cr 
\tablerule
\omit&height2pt&\omit&&\omit&\cr 
&& $\epsilon_2$ && $\frac{(1-2D)}{(D-1)}$ & \cr
\omit&height2pt&\omit&&\omit&\cr 
\tablerule
\omit&height2pt&\omit&&\omit&\cr 
&& $\zeta$ && $0$ & \cr
\omit&height2pt&\omit&&\omit&\cr 
\tablerule}}

\caption{Local Normal Contributions from Table~\ref{DivTerms}.}

\label{DivNew}

\end{table}

\begin{table}

\vbox{\tabskip=0pt \offinterlineskip
\def\tablerule{\noalign{\hrule}}
\halign to390pt {\strut#& \vrule#\tabskip=1em plus2em&
\hfil#\hfil& \vrule#& \hfil#\hfil& \vrule#\tabskip=0pt\cr
\tablerule \omit&height4pt&\omit&&\omit&\cr &&$\!\!\!\!{\rm
External \; Operator}\!\!\!\!$ && ${\rm Coefficient\ of}\;\;
\frac{i \kappa^2 H^4}{(4\pi)^4}$ & \cr
\omit&height4pt&\omit&&\omit&\cr 
\tablerule
\omit&height2pt&\omit&&\omit&\cr 
&& $\alpha$ && $\frac{\square}{H^2} [- \frac{\ln{x}}{3 x}]
+ \frac{2 \ln{x}}{3 x} - \frac1{3 x}$ & \cr
\omit&height2pt&\omit&&\omit&\cr 
\tablerule
\omit&height2pt&\omit&&\omit&\cr 
&& $\beta$ && $\frac{\square^2}{H^4} [\frac{\ln{x}}{6 x}]
+ \frac{\square}{H^2} [-\frac{\ln{x}}{3 x} + \frac1{6 x}]$ & \cr
\omit&height2pt&\omit&&\omit&\cr 
\tablerule
\omit&height2pt&\omit&&\omit&\cr 
&& $\gamma_1$ && $\frac{\square^2}{H^4} [\frac{2 \ln{x}}{x}]
+ \frac{\square}{H^2} [ -\frac{6 \ln{x}}{x} + \frac{2}{x}]
+ \frac{4 \ln{x}}{x} - \frac{2}{x}$ & \cr
\omit&height2pt&\omit&&\omit&\cr 
\tablerule
\omit&height2pt&\omit&&\omit&\cr 
&& $\gamma_2$ && $\frac{\square}{H^2} [ -\frac{2 \ln{x}}{3 x} ]
+ \frac{4 \ln{x}}{3 x} - \frac{2}{3 x}$ & \cr
\omit&height2pt&\omit&&\omit&\cr 
\tablerule
\omit&height2pt&\omit&&\omit&\cr 
&& $\delta$ && $0$ & \cr
\omit&height2pt&\omit&&\omit&\cr 
\tablerule
\omit&height2pt&\omit&&\omit&\cr 
&& $\epsilon_1$ && $\frac{\square}{H^2} [\frac{\ln{x}}{3 x} ]
- \frac{2 \ln{x}}{3 x} + \frac{1}{3 x}$ & \cr
\omit&height2pt&\omit&&\omit&\cr 
\tablerule
\omit&height2pt&\omit&&\omit&\cr 
&& $\epsilon_2$ && $\frac{\square}{H^2} [ \frac{7 \ln{x}}{6 x}]
- \frac{7 \ln{x}}{3 x} + \frac{7}{6 x}$ & \cr
\omit&height2pt&\omit&&\omit&\cr 
\tablerule
\omit&height2pt&\omit&&\omit&\cr 
&& $\zeta$ && $0$ & \cr
\omit&height2pt&\omit&&\omit&\cr 
\tablerule}}

\caption{Finite Normal Contributions from Table~\ref{DivTerms} with
$x = \frac{y}4$.}

\label{DivFin}

\end{table}

The next step is to reexpress the local terms of Table~\ref{DivNew} as
local counterterms. This is done by using the delta function to convert 
all factors of $a'$ from the External Operators into factors
of $a$, and then passing all factors of $a$ to the left. In most cases
this is straightforward but $\beta \frac{\square}{H^2}$ and $\beta$
require the following identities:
\begin{eqnarray}
\lefteqn{(aa')^{D-1} (a^2+a'^{2}) \square^2 \Bigl[a^{-D} 
\delta^D(x \!-\!x')\Bigr] = \Bigl[ 2a^D \square^2 -
12 H^2 a^D \square } \nonumber \\
& & \hspace{3cm} + 8 a^{D-2} H^2 \nabla^2 + 2 (D^2 \!-\! 2D \!+\! 2)
H^4a^D \Bigr]\delta^D(x \!-\! x') \; , \qquad \label{BoxID} \\
\lefteqn{(a a')^{D-1} (a^2 \!+\! a^{\prime 2}) H^2 \square \Bigl[a^{-D} 
\delta^D(x \!-\! x') \Bigr] = 2 a^D (H^2 \square - H^4) \delta^D(x \!-\! x') 
\; . }
\end{eqnarray}
Our results for the three possible counterterms (\ref{alp1}), 
(\ref{alp2}) and (\ref{alp3}) are reported in Table~\ref{NorLoc}. Note 
that the contribution to (\ref{alp1}) vanishes, as it must because this
counterterm happens to be zero in flat space. 

\begin{table}

\vbox{\tabskip=0pt \offinterlineskip
\def\tablerule{\noalign{\hrule}}
\halign to390pt {\strut#& \vrule#\tabskip=1em plus2em&
\hfil#\hfil& \vrule#& \hfil#\hfil& \vrule#& \hfil#\hfil&
\vrule#& \hfil#\hfil& \vrule#\tabskip=0pt\cr 
\tablerule
\omit&height4pt&\omit&&\omit&&\omit&&\omit&\cr
&& $\!\!\!\!\!{\rm From}\!\!\!\!\!$ && $\!\!\!\! a^D \square^2 
\delta^D(x \!-\! x') \!\!\!\!$ && $\!\!\!\! a^D H^2 \square 
\delta^D(x \!-\! x') \!\!\!\!$ && $\!\!\!\!\! a^{D-2} H^2 \nabla^2 
\delta^D(x \!-\! x') \!\!\!\!\!$ & \cr
\omit&height4pt&\omit&&\omit&&\omit&&\omit&\cr
\tablerule
\omit&height2pt&\omit&&\omit&&\omit&&\omit&\cr 
&& $\alpha$ && $\frac{D}{(D-1)(D-2)}$ && $0$ && $0$ & \cr
\omit&height2pt&\omit&&\omit&&\omit&&\omit&\cr 
\tablerule
\omit&height2pt&\omit&&\omit&&\omit&&\omit&\cr 
&& $\beta \frac{\square}{H^2}$ && $-\frac{D}{(D-1)(D-2)}$ 
&& $\frac{6 D}{(D-1)(D-2)}$ && $-\frac{4 D}{(D-1)(D-2)}$ & \cr
\omit&height2pt&\omit&&\omit&&\omit&&\omit&\cr 
\tablerule
\omit&height2pt&\omit&&\omit&&\omit&&\omit&\cr 
&& $\beta$ && $0$ && $-\frac{(D+2)(D-4)}{2}$ && $0$ & \cr
\omit&height2pt&\omit&&\omit&&\omit&&\omit&\cr 
\tablerule
\omit&height2pt&\omit&&\omit&&\omit&&\omit&\cr 
&& $\gamma_1 \frac{\square}{H^2}$ && $0$ && $-\frac{D^2}{2(D-2)}$
&& $0$ & \cr
\omit&height2pt&\omit&&\omit&&\omit&&\omit&\cr 
\tablerule
\omit&height2pt&\omit&&\omit&&\omit&&\omit&\cr 
&& $\epsilon_1$ && $0$ && $0$ && $\frac{(D^2 - 6 D + 4)}{2 (D-1)}$ & \cr
\omit&height2pt&\omit&&\omit&&\omit&&\omit&\cr 
\tablerule
\omit&height2pt&\omit&&\omit&&\omit&&\omit&\cr 
&& $\epsilon_2$ && $0$ && $0$ && $\frac{(2 - 4 D)}{(D-1)}$ & \cr
\omit&height2pt&\omit&&\omit&&\omit&&\omit&\cr 
\tablerule
\omit&height2pt&\omit&&\omit&&\omit&&\omit&\cr 
\tablerule
\omit&height2pt&\omit&&\omit&&\omit&&\omit&\cr 
&& ${\rm Total}$ && $0$ && $\frac{(D-4)(-D^3 + D - 4)}{2(D-1)(D-2)}$ 
&& $\frac{(D^3 - 16 D^2 + 28 D - 16)}{2 (D-1) (D-2)}$ & \cr
\omit&height2pt&\omit&&\omit&&\omit&&\omit&\cr 
\tablerule}}

\caption{Normal Contributions to Counterterms from Table~\ref{DivNew}.
All terms are multiplied by $\frac{i \kappa^2 H^{D-4}}{(4\pi)^{\frac{D}2}}
\frac{\Gamma(\frac{D}2)}{(D-3)(D-4)}$.}

\label{NorLoc}

\end{table}

Another important consistency check comes from the local terms proportional 
to $i \kappa^2 H^4 a^D \delta^D(x \!-\! x')$, which are reported in 
Table~\ref{OthLoc}. Recall that a counterterm of this form is forbidden by 
the symmetry $\phi \rightarrow \phi + {\rm const}$ of the bare Lagrangian 
(\ref{Lag}). Although three of the four contributions to Table~\ref{OthLoc}
diverge, their sum is finite for $D=4$. It doesn't vanish because the 
$A$-type propagator equation implies,
\begin{equation}
i a^D \delta^D(x \!-\! x') = (a a')^D \Bigl\{ \square A(y) - 
(D\!-\! 1) k H^2 \Bigr\} \; . \label{proprel}
\end{equation}
Because the total for Table~\ref{OthLoc} is finite one can take $D=4$ and
then use (\ref{proprel}) to subsume the result into finite, nonlocal terms
of the same form as have already been reported in Table~\ref{Finite Terms},
\begin{eqnarray}
{\rm Table~\ref{OthLoc}} & \!\!\!=\!\!\! & \frac{i \kappa^2 H^{D-4}}{
(4 \pi)^{\frac{D}2}} \frac{(-D^4 \!+\! 5 D^3 \!-\! 16 D \!+\! 16) 
\Gamma(\frac{D}2)}{4 (D\!-\!2) (D\!-\!3)} \times a^D H^4 \delta^D(x \!-\! x') 
\; , \qquad \\
& \!\!\!\longrightarrow \!\!\!& \frac{\kappa^2 H^4}{8 \pi^2} \times i a^4 
\delta^4(x \!-\! x') \; , \\
& \!\!\!=\!\!\! & \frac{\kappa^2 H^4}{(4 \pi)^4} \Biggl\{ (a a')^4 H^2 
\square \Bigl[ 2 \times \frac{4}{y} - 4 \ln\Bigl(\frac{y}4\Bigr)\Bigr] 
- (a a')^4 H^4 \times 12 \Biggr\} . \qquad \label{loccon}
\end{eqnarray}
Table~\ref{FinNew} includes this with the similarly finite results of 
Tables~\ref{LogNew}, \ref{Finite Terms} and \ref{DivFin}.

\begin{table}

\vbox{\tabskip=0pt \offinterlineskip
\def\tablerule{\noalign{\hrule}}
\halign to390pt {\strut#& \vrule#\tabskip=1em plus2em&
\hfil#\hfil& \vrule#& \hfil#\hfil& \vrule#\tabskip=0pt\cr 
\tablerule
\omit&height4pt&\omit&&\omit&\cr
&& $\!\!\!\!\!{\rm Contrib.\ from}\!\!\!\!\!$ && $\!\!\!\!\! {\rm Coef.\ of} 
\; \frac{i \kappa^2 H^{D-4}}{(4 \pi)^{\frac{D}2}} \frac{\Gamma(\frac{D}2)}{
(D-3)(D-4)} \times a^D H^4 \delta^D(x \!-\! x') \!\!\!\!\!$ & \cr
\omit&height4pt&\omit&&\omit&\cr
\tablerule
\omit&height2pt&\omit&&\omit&\cr 
&& $\beta \frac{\square}{H^2}$ && $-\frac{D (D^2 -2 D +2)}{(D-1)(D-2)}$ & \cr
\omit&height2pt&\omit&&\omit&\cr 
\tablerule
\omit&height2pt&\omit&&\omit&\cr 
&& $\beta$ && $\frac{(D+2)(D-4)}{2}$ & \cr
\omit&height2pt&\omit&&\omit&\cr 
\tablerule
\omit&height2pt&\omit&&\omit&\cr 
&& $\gamma_1$ && $-\frac{(D^4 - 5 D^3 + 16 D - 16)}{4}$ & \cr
\omit&height2pt&\omit&&\omit&\cr 
\tablerule
\omit&height2pt&\omit&&\omit&\cr 
&& $\gamma_2$ && $\frac{(D-2)(D^3 - 3 D^2 - 4 D + 8)}{2 (D-1)}$ & \cr
\omit&height2pt&\omit&&\omit&\cr 
\tablerule
\omit&height2pt&\omit&&\omit&\cr 
\tablerule
\omit&height2pt&\omit&&\omit&\cr 
&& ${\rm Total}$ && $\frac{(D-4)(-D^4 + 5 D^3 - 16 D + 16)}{4 (D-2)}$ & \cr
\omit&height2pt&\omit&&\omit&\cr 
\tablerule}}

\caption{Other Local Normal Contributions from Table~\ref{DivNew}.}

\label{OthLoc}

\end{table}

\begin{table}

\vbox{\tabskip=0pt \offinterlineskip
\def\tablerule{\noalign{\hrule}}
\halign to390pt {\strut#& \vrule#\tabskip=1em plus2em&
\hfil#\hfil& \vrule#& \hfil#\hfil& \vrule#\tabskip=0pt\cr
\tablerule \omit&height4pt&\omit&&\omit&\cr &&$\!\!\!\!{\rm
External \; Operator}\!\!\!\!$ && ${\rm Coefficient\ of}\;
\frac{ \kappa^2 H^4}{(4\pi)^4}$ & \cr
\omit&height4pt&\omit&&\omit&\cr 
\tablerule
\omit&height2pt&\omit&&\omit&\cr 
&& $(aa')^4 \square^3/H^2$ && $-\frac{\ln{x}}{3x}$ & \cr 
\omit&height2pt&\omit&&\omit&\cr
\tablerule 
\omit&height2pt&\omit&&\omit&\cr 
&& $(aa')^4 \square^2$ && $\frac{26\ln{x}}{3x} +\frac{38}{3x}
-6\ln^2{x} -18\ln{x}$ & \cr
\omit&height2pt&\omit&&\omit&\cr 
\tablerule
\omit&height2pt&\omit&&\omit&\cr 
&& $(aa')^4 H^2\square$ && $-\frac{6\ln{x}}{x} + \frac{4}{x} -4 \ln{x}$ & \cr
\omit&height2pt&\omit&&\omit&\cr 
\tablerule
\omit&height2pt&\omit&&\omit&\cr 
&& $(aa')^4 H^4$ && $\frac{4\ln{x}}{x}+\frac{18}{x}-120 -108\ln{x}$ & \cr
\omit&height2pt&\omit&&\omit&\cr 
\tablerule
\omit&height2pt&\omit&&\omit&\cr 
&& $(aa')^3(a^2+a'^2) \square^3/H^2$ && $\frac{\ln{x}}{6x}$ & \cr
\omit&height2pt&\omit&&\omit&\cr 
\tablerule
\omit&height2pt&\omit&&\omit&\cr 
&& $(aa')^3(a^2+a'^2) \square^2$ && $-\frac{\ln{x}}{3x}+\frac{1}{6x}$ & \cr
\omit&height2pt&\omit&&\omit&\cr 
\tablerule
\omit&height2pt&\omit&&\omit&\cr 
&& $(aa')^3(a^2+a'^2)H^2 \square$ && 
$-\frac{2\ln{x}}{3x}+\frac{5}{x}-18\ln{x}$ & \cr
\omit&height2pt&\omit&&\omit&\cr 
\tablerule
\omit&height2pt&\omit&&\omit&\cr 
&& $(aa')^3(a^2+a'^2)H^4$ && $\frac{4\ln{x}}{3x}-\frac{32}{3x}-54$ & \cr
\omit&height2pt&\omit&&\omit&\cr 
\tablerule
\omit&height2pt&\omit&&\omit&\cr 
&& $(aa')^3H^2\nabla^2$ && $-\frac{2\ln{x}}{3x}-\frac{16}{x} 
+84\ln{x}-48x\ln{x} + 96 x$ & \cr
\omit&height2pt&\omit&&\omit&\cr 
\tablerule
\omit&height2pt&\omit&&\omit&\cr 
&& $(aa')^3 \nabla^2 \square$ && $\frac{\ln{x}}{3x}$ & \cr 
\omit&height2pt&\omit&&\omit&\cr
\tablerule 
\omit&height2pt&\omit&&\omit&\cr 
&& $(aa')^2(a^2+a'^2)H^2\nabla^2$ && $-\frac{7\ln{x}}{3x} + \frac{11}{2x}
-12\ln{x}+48x\ln{x} + 12 x$ & \cr
\omit&height2pt&\omit&&\omit&\cr 
\tablerule
\omit&height2pt&\omit&&\omit&\cr 
&& $(aa')^2(a^2+a'^2)\nabla^2\square$ && $\!\!\!\!\!-\frac{17\ln{x}}{6x} 
- \frac{49}{6x} + 4 \ln^2{x} + 10 \ln{x} + 12 x \ln{x}\!\!\!\!\!$ & \cr 
\omit&height2pt&\omit&&\omit&\cr 
\tablerule
\omit&height2pt&\omit&&\omit&\cr 
&& $(aa')^2\nabla^4$ && $\frac{10}{3}\ln{x} - 24 x \ln{x}
+ 24 x^2 \ln{x} - 36 x^2$ & \cr
\omit&height2pt&\omit&&\omit&\cr 
\tablerule}}

\caption{All Finite Nonlocal Contributions with $x \equiv \frac{y}{4}$,
where $y(x;x')$ is defined in equation (\ref{ydef}).}

\label{FinNew}

\end{table}

Our final result for the regulated but unrenormalized, one loop
self-mass-squared derives from combining expressions (\ref{fin4pt}), 
(\ref{fin3loc}), and the local parts of (\ref{fin3log}), with Tables
\ref{NorLoc} and \ref{FinNew}. It takes the form,
\begin{equation}
-i M^2_{\rm reg}(x;x') = i \kappa^2 a^D \Bigl(\beta_1 \square^2
+ \beta_2 \square + \beta_3 \frac{\nabla^2}{a^2}\Bigr) \delta^D(x \!-\! x')
+ {\rm Table~\ref{FinNew}} + O(D\!-\!4) \; .
\end{equation}
The coefficients $\beta_i$ are,
\begin{eqnarray}
\beta_1 \!\!\!\!&=&\!\!\!\! 0 \; , \\
\beta_2 \!\!\!\!&=&\!\!\!\! \frac{H^{D-4}}{(4\pi)^{\frac{D}2}} \Biggl\{ 
\frac{(-D^3 \!+\! D \!-\! 4) \Gamma(\frac{D}2 \!-\! 1)}{4 (D\!-\!1) (D\!-\!3)}
- \frac{(D\!+\!1) (D\!-\!4) \Gamma(D) \pi \cot(\frac{\pi}2 D)}{4 (D\!-\!3) 
\Gamma(\frac{D}2)} \Biggr\} , \qquad \\
\!\!\!\!& = &\!\!\!\! \frac{H^{D-4}}{(4\pi)^{\frac{D}2}} \Biggl\{-\frac{61}3 
+ O(D\!-\!4) \Biggr\} \; , \qquad \\
\beta_3 \!\!\!\!&=&\!\!\!\! \frac{H^{D-4}}{(4\pi)^{\frac{D}2}} \Biggl\{ 
\frac{(D^3 \!-\! 16 D^2 \!+\! 28 D \!-\! 16) \Gamma(\frac{D}2)}{2 (D\!-\!1)
(D\!-\!2) (D\!-\!3) (D\!-\!4)} \nonumber \\
& & \hspace{4cm} + \frac{(D^2 \!-\! 4 D \!+\! 1) \Gamma(D \!-\!1) \pi 
\cot(\frac{\pi}2 D)}{(D \!-\! 3) \Gamma(\frac{D}2)} - 3 \Biggr\} , \qquad \\
\!\!\!\!& = &\!\!\!\! \frac{H^{D-4}}{(4\pi)^{\frac{D}2}} \Biggl\{- 
\frac4{D\!-\!4} + \frac{58}3 + 2 \gamma + O(D\!-\!4) \Biggr\} \; .
\end{eqnarray}
(Here $\gamma \sim .577215$ is Euler's constant.)
The obvious renormalization convention is to choose each of the three 
$\alpha_i$'s to absorb the corresponding $\beta_i$, leaving an arbitrary
finite term $\Delta \alpha_i$,
\begin{equation}
\alpha_i = -\beta_i + \Delta \alpha_i \; .
\end{equation}
We can now take the unregulated limit ($D=4$) to obtain the final
renormalized result,
\begin{equation}
-i M^2_{\rm ren}(x;x') = i \kappa^2 a^4 \Bigl(\Delta \alpha_1 \square^2
+ \Delta \alpha_2 \square + \Delta \alpha_3 \frac{\nabla^2}{a^2}\Bigr) 
\delta^4(x \!-\! x') + {\rm Table~\ref{FinNew}} \; . \label{final}
\end{equation}

\section{Discussion}

We have computed one loop quantum gravitational corrections to the 
scalar self-mass-squared on a locally de Sitter background. The
computation was done using dimensional regularization and renormalized
by subtracting the three possible BPHZ counterterms. Because our gauge 
condition (\ref{GF}) breaks de Sitter invariance, one of these counterterms
is noninvariant. Our final result, expression (\ref{final}), consists 
of arbitrary finite contributions from the three counterterms plus
the nonlocal contributions given in Table~\ref{FinNew}.

The point of this exercise is to discover whether or not the inflationary
production of gravitons has a significant effect upon minimally coupled
scalars as it does on fermions \cite{MW2}. We will check this in a 
subsequent paper \cite{subs} by computing one loop corrections to the 
scalar mode functions using the effective field equation,
\begin{equation}
\partial_{\mu} \Bigl( \sqrt{-\widehat{g}} \, \widehat{g}^{\mu\nu} 
\partial_{\nu} \Phi(x) \Bigr) - \int d^4x' M^2_{\rm ren}(x;x') 
\Phi(x') = 0 \; .
\end{equation}
Similar studies have already probed the effects of scalar 
self-inter\-ac\-tions \cite{BOW,KO}, fermions \cite{DW} and photons 
\cite{KW2}, but none has so far considered the effects of gravitons. 
Although our scalar is a spectator to $\Lambda$-driven inflation, the 
near flatness of inflaton potentials suggests that the result we shall
obtain may apply as well to the inflaton of scalar-driven inflation.

A significant difference between this and previous scalar studies 
\cite{BOW,KO,DW,KW2} is that quantum gravity is not renormalizable.
Although we could absorb divergences with quartic, BPHZ counterterms,
no physical principle fixes the finite coefficients $\Delta \alpha_i$ of
these counterterms. That ambiguity is one way of expressing the problem 
of quantum gravity. However, a little thought reveals that we will be
able to get unambiguous results for late time corrections to the mode
functions. The reason is that the scalar d'Alembertian annihilates the 
tree order mode solution,
\begin{equation}
\Phi_0(x;\vec{k}) = u(\eta,k) e^{i \vec{k} \cdot \vec{x}}
\quad {\rm where} \quad u(\eta,k) = \frac{H}{\sqrt{2 k^3}} \Bigl[1 -
\frac{i k}{H a} \Bigr] \exp\Bigl[\frac{i k}{H a}\Bigr] \; .
\end{equation}
Hence only the third counterterm makes a nonzero contribution, and its 
effect rapidly redshifts away,
\begin{eqnarray}
\lefteqn{\int d^4x' \, \kappa^2 a^4 \Bigl(\Delta \alpha_1 \square^2 
+ \Delta \alpha_2 \square + \Delta \alpha_3 \frac{\nabla^2}{a^2} \Bigr) 
\delta^4(x \!-\! x') \times \Phi_0(x';\vec{k}) } \nonumber \\
& & \hspace{6.5cm} = -\kappa^2 a^4 \times \Delta \alpha_3 \frac{k^2}{a^2} \,
\Phi_0(x;\vec{k}) \; . \qquad
\end{eqnarray}

It is instructive to compare our de Sitter background result 
(\ref{final}) with its flat space analogue. In the flat space limit of
$H \rightarrow 0$ with fixed comoving time, the scalar and graviton 
propagators become,
\begin{eqnarray}
i\Delta_A^{\rm flat}(x;x') & = & \frac{\Gamma(\frac{D}2 \!-\! 1)}{4
\pi^{\frac{D}2}} \, \frac1{\Delta x^{D-2}} \; , \\
i\Bigl[\mbox{}_{\alpha\beta} \Delta^{\rm flat}_{\rho\sigma}\Bigr](x;x')
& = & \Bigl[2 \eta_{\alpha (\rho} \eta_{\sigma) \beta} \!-\! 
\frac2{D\!-\!2} \eta_{\alpha\beta} \eta_{\rho\sigma} \Bigr] 
\frac{\Gamma(\frac{D}2 \!-\! 1)}{4 \pi^{\frac{D}2}} \, 
\frac1{\Delta x^{D-2}} \; .
\end{eqnarray}
Here $\Delta x^2$ is the Poincar\'e length function analogous to $y(x;x')$,
\begin{equation}
\Delta x^2(x;x') \equiv \Vert \vec{x} \!-\! \vec{x}' \Vert^2 -
\Bigl(\vert \eta \!-\! \eta' \vert - i \delta \Bigr)^2 \; .
\end{equation}
Two features of the flat space propagators deserve comment:
\begin{enumerate}
\item{Both propagators are manifestly Poincar\'e invariant; and}
\item{The coincidence limits of both propagators vanish in dimensional
regularization.}
\end{enumerate}
We exploited the first property to drop all but the final noninvariant 
counterterm on the list (\ref{noninv}). And the second property explains 
why our 4-point contribution (\ref{fin4pt}) vanishes in the flat space 
limit,
\begin{eqnarray}
\lefteqn{-i M^2_{{\rm flat} \atop {\rm 4pt}}(x;x') = -\frac{i \kappa^2}8
\partial^{\mu} \Biggl\{ i\Bigl[\mbox{}^{\alpha}_{~ \alpha} \Delta^{\rm flat}_{
\rho\sigma}\Bigr]\!(x;x) \, \eta^{\rho\sigma} \partial_{\mu} \delta^D(x\!-\!x') 
\Biggr\} } \nonumber \\
& & \hspace{-.7cm} + \frac{i \kappa^2}4 \partial^{\mu} \Biggl\{\! i\Bigl[
\mbox{}^{\alpha\beta} \Delta^{\rm flat}_{\alpha\beta}\Bigr]\!(x;x) \,
\partial_{\mu} \delta^D(x\!-\!x') \!\Biggr\} \!+\! \frac{i \kappa^2}2 
\partial^{\rho} \Biggl\{ \!i\Bigl[\mbox{}^{\alpha}_{~\alpha} \Delta^{\rm 
flat}_{\rho\sigma} \Bigr]\!(x;x) \, \partial^{\sigma} \delta^D(x\!-\!x') \!
\Biggr\} \nonumber \\
& & \hspace{4cm} -i \kappa^2 \partial_{\alpha} \Biggl\{ i\Bigl[\mbox{}^{
\alpha\rho} \Delta^{\rm flat}_{\rho\sigma} \Bigr]\!(x;x) \, \partial^{\sigma} 
\delta^D(x\!-\!x') \Biggr\} = 0 \; . \qquad
\end{eqnarray}

An only slightly less trivial computation reveals that the flat space
limit of our total 3-point contribution should also vanish,
\begin{eqnarray}
\lefteqn{-iM^2_{{\rm flat} \atop {\rm 3pt}}(x;x') = -\kappa^2 
\partial^{\alpha} \partial^{\prime\rho} \Biggl\{ i\Bigl[\mbox{}_{\alpha\beta} 
\Delta^{\rm flat}_{\rho\sigma}\Bigr]\!(x;x') \partial^{\beta} 
\partial^{\prime\sigma} i\Delta^{\rm flat}_A(x;x') \Biggr\} } \nonumber \\
& & \hspace{.4cm} + \frac{\kappa^2}2 \partial^{\mu} \partial^{\prime\rho} 
\Biggl\{ i\Bigl[\mbox{}^{\alpha}_{~\alpha} \Delta^{\rm flat}_{\rho\sigma} 
\Bigr] \partial_{\mu} \partial^{\prime\sigma} i\Delta_A^{\rm flat} \Biggr\}\!
+ \frac{\kappa^2}2 \partial^{\alpha} \partial^{\prime \nu} \Biggl\{ 
i\Bigl[\mbox{}_{\alpha\beta} \Delta^{\rm flat}_{\rho\sigma}\Bigr] \, 
\eta^{\rho\sigma} \partial^{\beta} \partial_{\nu}' i\Delta_A^{\rm flat}
\Biggr\} \nonumber \\
& & \hspace{3cm} - \frac{\kappa^2}4 \partial^{\mu} \partial^{\prime \nu}
\Biggl\{ i\Bigl[\mbox{}^{\alpha}_{~\alpha} \Delta^{\rm flat}_{\rho\sigma} 
\Bigr]\!(x;x') \, \eta^{\rho\sigma} \partial_{\mu} \partial_{\nu}' 
i\Delta_A^{\rm flat}(x;x') \Biggr\} \; , \qquad \\
& & = -\frac{\kappa^2 \Gamma^2(\frac{D}2 \!-\! 1)}{16 \pi^D} \partial^{\alpha}
\partial^{\rho} \Biggl\{ \frac1{\Delta x^{D-2}} \, \partial^{\beta} 
\partial^{\sigma} \frac1{\Delta x^{D-2}} \Biggr\} \Bigl[ 2 \eta_{\alpha (\rho}
\eta_{\sigma) \beta} - \eta_{\alpha\beta} \eta_{\rho\sigma}\Bigr] \; , \qquad\\
& & = -\frac{\kappa^2 \Gamma^2(\frac{D}2 \!-\! 1)}{16 \pi^D} \partial^2
\Biggl\{ \frac1{\Delta x^{D-2}} \, \partial^2 \frac1{\Delta x^{D-2}} \Biggr\} 
\; , \qquad \\
& & = 0 \; .
\end{eqnarray}
This result has a number of consequences:
\begin{enumerate}
\item{It explains why the highest dimension counterterm (\ref{alp1}) fails
to appear;}
\item{It explains why all the entries of Table~\ref{FinNew} vanish in the
flat space limit except the first line,
\begin{eqnarray}
\lefteqn{\frac{\kappa^2 H^2}{(4 \pi)^4} \, (a a')^4 \square^3 \Biggl\{ 
-\frac13 \times \frac{4}{y} \ln\Bigl(\frac{y}{4}\Bigr) \Biggr\} } \nonumber \\
& & \hspace{3cm} \longrightarrow -\frac13 \times \frac{\kappa^2}{(4\pi)^4} \, 
\partial^6 \Biggl\{ \frac{4}{\Delta x^2} \ln\Bigl(\frac14 H^2 \Delta x^2\Bigr) 
\Biggr\} \; , \qquad \label{tab1}
\end{eqnarray}
and the fifth line,
\begin{eqnarray}
\lefteqn{\frac{\kappa^2 H^2}{(4 \pi)^4} \, (a a')^3 (a^2 + a^{\prime 2}) 
\square^3 \Biggl\{ \frac16 \times \frac{4}{y} \ln\Bigl(\frac{y}{4}\Bigr) 
\Biggr\} } \nonumber \\
& & \hspace{3cm} \longrightarrow +\frac13 \times \frac{\kappa^2}{(4\pi)^4} \, 
\partial^6 \Biggl\{ \frac{4}{\Delta x^2} \ln\Bigl(\frac14 H^2 \Delta x^2
\Bigr) \Biggr\} \; ; \qquad \label{tab5}
\end{eqnarray}}
\item{It explains why (\ref{tab1}) and (\ref{tab5}) cancel in the flat
space limit; and}
\item{It means that any physical effect we find must derive entirely from 
the nonzero Hubble constant.}
\end{enumerate}

This is the right point to comment on accuracy. This has been a long and
tedious computation, involving the combination of many distinct pieces. It
is significant when these pieces join together to produce results that can
be checked independently, such as the vanishing of the flat space limit. 
One sees that in the way the $\alpha$ and $\beta \frac{\square}{H^2}$
contributions to the $\alpha_1$ counterterm cancel in Table~\ref{NorLoc}.
Another example is the way three individually divergent terms combine in
Table~\ref{OthLoc} to produce a finite result for a counterterm that is
forbidden by the shift symmetry of the bare Lagrangian (\ref{Lag}). 

Although the $\alpha_1$ counterterm had to vanish by the flat space
limit, we do not yet understand why the coefficient the $\alpha_2$ 
counterterm is finite. The contribution of this term to the scalar
self-mass-squared vanishes in flat space, but it would seem to affect 
the $\phi + h \rightarrow \phi + h$ scattering amplitude. The
divergences on this were explored in the classic paper of 't Hooft 
and Veltman \cite{HV}. Unfortunately, their on-shell analysis makes no
distinction between $R (\partial \phi)^2$ --- which we have --- and 
$(\partial \phi)^4$ --- which we do not have.

Finally, we should comment on what subset of the full de Sitter group is 
respected by our result (\ref{final}). Recall that our gauge fixing
term breaks spatial special conformal transformation (\ref{sscx}).
This is why the noninvariant counterterm (\ref{alp3}) occurs. It is also
responsible for the noninvariant factors of $\nabla^2$ and $a^2 + 
a^{\prime 2}$ in Table~\ref{FinNew}. Because these breakings derive
entirely from the gauge condition, we expect them to have no physical
consequence.

The graviton and scalar propagators also break the dilatation symmetry 
(\ref{dilx}). Unlike the violation of spatial special conformal 
transformations, the breaking of dilatation invariance is a physical 
manifestation of inflationary particle production and can have important
consequences. Dilatation breaking comes in the $\ln(a a')$ term of the 
$A$-type propagator (\ref{A}). These logarithms were responsible for 
the secular growth that was found in the fermion field strength 
\cite{MW2}, so one might expect them to drive any effect on scalars 
as well. However, it turns out that the factors of $\ln(a a')$ all
drop out. For the scalar propagator this is a trivial consequence 
of the fact that it always carries one primed and one un-primed 
derivative. Logarithms from the graviton propagator do appear in the 
4-point contributions (\ref{fin4pt}), and in the 3-point logarithm 
contributions (\ref{fin3log}). But all factors of $\ln(a a')$ cancel
in the final result (\ref{final}), which turns out to respect
dilatation invariance. Because of this we suspect that there will be
no significant late time corrections to the mode functions at one
loop order.

\vskip .5cm

\centerline{\bf Acknowledgements}

This work was partially supported by NSF grant PHY-0653085 and by the
Institute for Fundamental Theory at the University of Florida.

\section{Appendix: Extracting Derivatives}

We group the various normal contributions into seven parts:
\begin{eqnarray}
P_1 & \equiv & \nabla \!\cdot\! \nabla' \Bigl[ (a a')^{D-2} A
\nabla \!\cdot\! \nabla' A\Bigr] \; , \\
P_2 & \equiv & \partial_i \partial_0' \Bigl[(a a')^{D-2} A
\partial_i \partial_0' A\Bigr] + \partial_0 \partial_i' \Bigl[(a
a')^{D-2}A
\partial_0 \partial_i' A\Bigr] \; , \\
P_3 & \equiv & \partial_0 \partial_0' \Bigl[(a a')^{D-2} A
\partial_0 \partial_0' A\Bigr] \; , \\
P_4 & \equiv & -\partial_0 \partial_0' \Bigl[(a a')^{D-2} B
\nabla \!\cdot\! \nabla' A\Bigr] \; , \\
P_5 & \equiv & -\partial_i \partial_0' \Bigl[(a a')^{D-2} B
\partial_0 \partial_i' A\Bigr] - \partial_0 \partial_i' \Bigl[(a
a')^{D-2}B
\partial_i \partial_0' A\Bigr] \; , \\
P_6 & \equiv & -\nabla \!\cdot\! \nabla' \Bigl[ (a a')^{D-2} B
\partial_0 \partial_0' A\Bigr] \; , \\
P_7 & \equiv & \partial_0 \partial_0' \Bigl[(a a')^{D-2} \Delta C
\partial_0 \partial_0' A\Bigr] \; .
\end{eqnarray}
In these definitions the exprssion ``$\partial_0 \partial_0' A(y)$''
means the naive derivative, {\it without} the delta function. Also
note that we have suppressed the unbiquitous factors of $\kappa^2$.

An important simplification in reducing $P_2$ is to achieve a
symmetric form which has no $\partial_0$. This can be done by
adding equations (\ref{ID11}) and (\ref{ID14}) and then using
equation (\ref{ID3}),
\begin{eqnarray}
P_2 & = & (-\delta + 2 \zeta) I^2[A A''] + (D\!-\!1) \epsilon_2
I^2[A^{\prime 2}] \nonumber \\
& & \hspace{3cm} + H (a a')^{D-2} (a \partial_0 \!+\! a'
\partial_0')
\nabla^2 I^2[A^{\prime 2}] \; , \\
& = & (-\delta + 2 \zeta) I^2[A A''] + (D\!-\!1) \epsilon_2
I^2[A^{\prime 2}]  \nonumber \\
& & \hspace{3cm} + \epsilon_3 y I[A^{\prime 2}] + (D\!-\!1)
\epsilon_1 I^2[A^{\prime 2}] - \frac{\zeta}2 I^3[A^{\prime 2}] \;
. \qquad \label{P2}
\end{eqnarray}

Another organizational point concerns removing factors of $y$ from
inside integrals. This is desirable because it reduces the number
of distinct integrals which appear. It can always be accomplished
by partial integration. We will illustrate using the function
acted upon by $-\epsilon_3$ in equation (\ref{line2}),
\begin{equation}
F(y) \equiv (D\!-\!2) I[A'B \!+\! y A''B] + y A' B + y^2 A'' B - y
I[A'B'] \; .
\end{equation}
Note the relations,
\begin{eqnarray}
A' B + y A'' B & = & A' B + y \frac{\partial}{\partial y} I[A'' B] \; , \\
& = & \frac{\partial}{\partial y} \Bigl\{y I[A'' B] \Bigr\} + A' B
-
I[A'' B] \; , \\
& = & \frac{\partial}{\partial y} \Bigl\{y I[A'' B]\Bigr\} + I[A'
B'] \; . \label{simp1}
\end{eqnarray}
We can therefore write,
\begin{equation}
F(y) = y^2 A'' B + (D\!-\!1) y I[A'' B] + (D\!-\!2) I^2[A' B'] \;
. \label{simp2}
\end{equation}

With (\ref{simp1}) and (\ref{simp2}) we can read off the following
result for $P_5$ from equation (\ref{line2}),
\begin{eqnarray}
P_5 & = & \delta I^2[B A''] - (D\!-\!1) \epsilon_1 I^2[B' A']
- (D\!-\!1) \epsilon_2 I^2[B' A'] \nonumber \\
& & \hspace{1cm} + \zeta \Bigl\{-2 I^2[B A''] + \frac12 I^3[B' A']
\Bigr\}
+ \epsilon_2 \Bigl\{y I[A'' B] + I^2[A' B']\Bigr\} \nonumber \\
& & \hspace{2.3cm} + \epsilon_3 \Bigl\{y^2 A'' B + (D\!-\!1) y
I[A'' B] + (D\!-\!2) I^2[A' B'] \Bigr\} . \qquad \label{P5}
\end{eqnarray}
Many terms involving $A$ in $P_2$ combine with cognate terms
involving $B$ in $P_5$ to produce the less singular propagator
function $\Delta B = B - A$. Summing expressions (\ref{P2}) and
(\ref{P5}) gives,
\begin{eqnarray}
P_{2+5} \!\!\!\!\!& = & \!\!\!\!\! \delta I^2[\Delta B A''] -
(D\!-\!1) \epsilon_1 I^2[\Delta B' A'] - (D\!-\!1) \epsilon_2
I^2[\Delta B' A']
+ \epsilon_3 y I[A^{\prime 2}] \nonumber \\
& & \hspace{1cm} + \zeta \Bigl\{-2 I^2[\Delta B A''] + \frac12
I^3[\Delta B' A'] \Bigr\} + \epsilon_2 \Bigl\{y I[A'' B] +
I^2[A' B']\Bigr\} \nonumber \\
& & \hspace{2cm} + \epsilon_3 \Bigl\{y^2 A'' B + (D\!-\!1) y
I[A'' B] + (D\!-\!2) I^2[A' B'] \Bigr\} \; . \qquad
\end{eqnarray}

In contradistinction to $P_2$ and $P_5$, the reduction of the
other parts is facilitated by further sub-division immediately
after employing identities (\ref{ID013}) and (\ref{ID016}),
\begin{eqnarray}
f(y) \partial_0 \partial_0' A(y) & = & \partial_0 \partial_0'
I^2[f A''] +
2 a a' H^2 I[f' A'] \nonumber \\
& & \hspace{-.5cm} - a a' H^2 \Bigl\{(D\!-\!1) + y \frac{\partial}{\partial y}
\Bigr\} I^2[f' A'] - \frac12 \nabla \!\cdot\! \nabla' I^3[f' A']
\; , \qquad \label{ID1a} \\
f(y) \nabla \!\cdot\! \nabla' A(y) & = & \nabla \!\cdot\! \nabla'
I^2[f A''] - 2 (D\!-\!1) a a' H^2 I[f' A] \; . \label{ID2b}
\end{eqnarray}
One employs (\ref{ID1a}) on $P_3$ (from which we can read off the
result for $P_7$) and $P_6$ to give the sub-parts,
\begin{eqnarray}
P_{3a} & \equiv & \partial_0 \partial_0' \Bigl[ (a a')^{D-2}
\partial_0
\partial_0' I^2[A A''] \Bigr] \; , \\
P_{3b} & \equiv & 2 \partial_0 \partial_0' \Bigl[ (a a')^{D-1} H^2
I[A^{\prime 2}] \Bigr] \; , \\
P_{3c} & \equiv & - \partial_0 \partial_0' \Bigl[ (a a')^{D-1} H^2
\Bigl\{(D\!-\!1) + y \frac{\partial}{\partial y}\Bigr\}
I^2[A^{\prime 2}]
\Bigr] \; , \qquad \\
P_{3d} & \equiv & -\frac12 \partial_0 \partial_0' \Bigl[ (a
a')^{D-2}
\nabla \!\cdot\! \nabla' I^3[A^{\prime 2}] \Bigr] \; , \\
P_{7a} & \equiv & \partial_0 \partial_0' \Bigl[ (a a')^{D-2}
\partial_0
\partial_0' I^2[{\Delta C} A''] \Bigr] \; , \\
P_{7b} & \equiv & 2 \partial_0 \partial_0' \Bigl[ (a a')^{D-1} H^2
I[A' {\Delta C}'] \Bigr] \; , \\
P_{7c} & \equiv & - \partial_0 \partial_0' \Bigl[ (a a')^{D-1} H^2
\Bigl\{(D\!-\!1) + y \frac{\partial}{\partial y}\Bigr\} I^2[A'
{\Delta C}']
\Bigr] \; , \qquad \\
P_{7d} & \equiv & -\frac12 \partial_0 \partial_0' \Bigl[ (a
a')^{D-2}
\nabla \!\cdot\! \nabla' I^3[A' {\Delta C}'] \Bigr] \; , \\
P_{6a} & \equiv & -\nabla \!\cdot\! \nabla' \Bigl[ (a a')^{D-2}
\partial_0
\partial_0' I^2[B A''] \Bigr] \; , \\
P_{6b} & \equiv & -2 \nabla \!\cdot\! \nabla' \Bigl[ (a a')^{D-1}
H^2
I[A' B'] \Bigr] \; , \\
P_{6c} & \equiv & \nabla \!\cdot\! \nabla' \Bigl[ (a a')^{D-1} H^2
\Bigl\{(D\!-\!1) + y \frac{\partial}{\partial y}\Bigr\} I^2[A' B']
\Bigr] \; , \qquad \\
P_{6d} & \equiv & \frac12 \nabla \!\cdot\! \nabla' \Bigl[ (a
a')^{D-2} \nabla \!\cdot\! \nabla' I^3[A' B'] \Bigr] \; .
\end{eqnarray}
Applying the second identity (\ref{ID2b}) to $P_1$ and $P_6$ gives
their sub-parts,
\begin{eqnarray}
P_{1a} & \equiv & \nabla \!\cdot\! \nabla' \Bigl[ (a a')^{D-2}
\nabla \!\cdot\! \nabla' I^2[A A''] \Bigr] \; , \\
P_{1b} & \equiv & -2 (D\!-\!1) \nabla \!\cdot\! \nabla' \Bigl[ (a
a')^{D-1}
H^2 I[A^{\prime 2}] \Bigr] \; , \\
P_{4a} & \equiv & -\partial_0 \partial_0' \Bigl[ (a a')^{D-2}
\nabla \!\cdot\! \nabla' I^2[B A''] \Bigr] \; , \\
P_{4b} & \equiv & 2 (D\!-\!1) \partial_0 \partial_0' \Bigl[ (a
a')^{D-1} H^2 I[A' B'] \Bigr] \; .
\end{eqnarray}

Of course there is no problem further reducing the spatial
derivatives. The following generic reductions serve to reduce
terms involving the operator $\partial_0 \partial_0'$,
\begin{eqnarray}
\lefteqn{\partial_0 \partial_0' \Bigl[ (a a')^{D-2} \partial_0
\partial_0'
f(y)\Bigr] = \Bigl\{ \alpha - \delta + \zeta\Bigr\} f(y) \; , } \\
\lefteqn{\partial_0 \partial_0' \Bigl[ (a a')^{D-1} H^2 f(y)\Bigr]
= \Biggr\{\frac{\beta}2 + \frac12 (D\!-\!1) (D\!-\!2) \gamma_1 +
\frac{\gamma_2}2 y \frac{\partial}{\partial y} } \nonumber \\
& & \hspace{2cm} + \frac{\gamma_3}2 \Bigl[(D\!-\!1) y
\frac{\partial}{\partial y} + y^2 \frac{\partial^2}{\partial
y^2}\Bigr]
+ \epsilon_1 \Bigl[-1 + \frac{D}4 I\Bigr] \Biggr\} f(y) \; , \qquad \\
\lefteqn{\partial_0 \partial_0' \Bigl[ (a a')^{D-2} \nabla
\!\cdot\! \nabla' f(y)\Bigr] = \Biggr\{-\frac{\delta}2 - \frac12
(D\!-\!2) (D\!-\!3) \epsilon_1
- \frac{\epsilon_2}2 y \frac{\partial}{\partial y} } \nonumber \\
& & \hspace{2cm} - \frac{\epsilon_3}2 \Bigl[(D\!-\!1) y
\frac{\partial}{\partial y} + y^2 \frac{\partial^2}{\partial
y^2}\Bigr]
+ \zeta \Bigl[1 - \frac14 (D\!-\!2) I\Bigr] \Biggr\} f(y) \; , \qquad \\
\lefteqn{\nabla \!\cdot\! \nabla' \Bigl[ (a a')^{D-2} \partial_0
\partial_0' f(y)\Bigr] = \Biggr\{-\frac{\delta}2 - \frac12
(D\!-\!1) (D\!-\!2) \epsilon_1
- \frac{\epsilon_2}2 y \frac{\partial}{\partial y} } \nonumber \\
& & \hspace{2cm} - \frac{\epsilon_3}2 \Bigl[(D\!-\!1) y
\frac{\partial}{\partial y} + y^2 \frac{\partial^2}{\partial
y^2}\Bigr] + \zeta \Bigl[1 + \frac14 (D\!-\!2) I\Bigr] \Biggr\}
f(y) \; . \qquad
\end{eqnarray}
Tables~\ref{alpha}-\ref{zeta} give the results for each of the ten 
External Operators.

\begin{table}

\vbox{\tabskip=0pt \offinterlineskip
\def\tablerule{\noalign{\hrule}}
\halign to390pt {\strut#& \vrule#\tabskip=1em plus2em&
\hfil#\hfil& \vrule#& \hfil#\hfil& \vrule#\tabskip=0pt\cr
\tablerule \omit&height4pt&\omit&&\omit&\cr &&$\!\!\!\!{\rm
Part}\!\!\!\!$ && $\!\!\!\! {\rm Contribution\ Acted\ upon\ by} \;
\alpha \!\!\!\!$ & \cr \omit&height4pt&\omit&&\omit&\cr \tablerule
\omit&height2pt&\omit&&\omit&\cr && $P_{3a}$ && $I^2[A A'']$ & \cr
\omit&height2pt&\omit&&\omit&\cr \tablerule
\omit&height2pt&\omit&&\omit&\cr && $P_{7a}$ && $I^2[\Delta C
A'']$ & \cr \omit&height2pt&\omit&&\omit&\cr \tablerule
\omit&height2pt&\omit&&\omit&\cr \tablerule
\omit&height2pt&\omit&&\omit&\cr && Total && $I^2[A A''] +
I^2[{\Delta C} A'']$ & \cr \omit&height2pt&\omit&&\omit&\cr
\tablerule}}

\caption{Contributions acted upon by $\alpha = (a a')^D
\square^2$.}

\label{alpha}

\end{table}

\begin{table}

\vbox{\tabskip=0pt \offinterlineskip
\def\tablerule{\noalign{\hrule}}
\halign to390pt {\strut#& \vrule#\tabskip=1em plus2em&
\hfil#\hfil& \vrule#& \hfil#\hfil& \vrule#\tabskip=0pt\cr
\tablerule \omit&height4pt&\omit&&\omit&\cr &&$\!\!\!\!{\rm
Part}\!\!\!\!$ && $\!\!\!\! {\rm Contribution\ Acted\ upon\ by} \;
\beta \!\!\!\!$ & \cr \omit&height4pt&\omit&&\omit&\cr \tablerule
\omit&height2pt&\omit&&\omit&\cr && $P_{3b}$ && $I[A^{\prime 2}]$
& \cr \omit&height2pt&\omit&&\omit&\cr \tablerule
\omit&height2pt&\omit&&\omit&\cr && $P_{3c}$ && $-\frac12 y
I[A^{\prime 2}] - (\frac{D-1}2) I^2[A^{\prime 2}]$ & \cr
\omit&height2pt&\omit&&\omit&\cr \tablerule
\omit&height2pt&\omit&&\omit&\cr && $P_{4b}$ && ${\scriptstyle
(D-1)} I[A' B']$ & \cr \omit&height2pt&\omit&&\omit&\cr \tablerule
\omit&height2pt&\omit&&\omit&\cr && $P_{7b}$ && $I[A' {\Delta
C}']$ & \cr \omit&height2pt&\omit&&\omit&\cr \tablerule
\omit&height2pt&\omit&&\omit&\cr && $P_{7c}$ && $-\frac12 y I[A'
{\Delta C}'] - (\frac{D-1}2) I^2[A' {\Delta C}']$ & \cr
\omit&height2pt&\omit&&\omit&\cr \tablerule
\omit&height2pt&\omit&&\omit&\cr \tablerule
\omit&height2pt&\omit&&\omit&\cr && Total && ${\scriptstyle D}
I[A' B'] - I[A' {\Delta B}'] -\frac12 y I[A^{\prime 2}] -
(\frac{D-1}2) I^2[A^{\prime 2}]$ & \cr
\omit&height2pt&\omit&&\omit&\cr && \omit && $+ I[A' {\Delta C}']
-\frac12 y I[A' {\Delta C}'] - (\frac{D-1}2) I^2[A' {\Delta C}']$
& \cr \omit&height2pt&\omit&&\omit&\cr \tablerule}}

\caption{Contributions acted upon by $\beta = (a a')^{D-2} (a^2 +
a^{\prime 2}) H^2 \square$.}

\label{beta}

\end{table}

\begin{table}

\vbox{\tabskip=0pt \offinterlineskip
\def\tablerule{\noalign{\hrule}}
\halign to390pt {\strut#& \vrule#\tabskip=1em plus2em&
\hfil#\hfil& \vrule#& \hfil#\hfil& \vrule#\tabskip=0pt\cr
\tablerule \omit&height4pt&\omit&&\omit&\cr &&$\!\!\!\!{\rm
Part}\!\!\!\!$ && $\!\!\!\! {\rm Contribution\ Acted\ upon\ by} \;
\gamma_1 \!\!\!\!$ & \cr \omit&height4pt&\omit&&\omit&\cr
\tablerule \omit&height2pt&\omit&&\omit&\cr && $P_{3b}$ &&
${\scriptstyle (D-1)(D-2)} I[A^{\prime 2}]$ & \cr
\omit&height2pt&\omit&&\omit&\cr \tablerule
\omit&height2pt&\omit&&\omit&\cr && $P_{3c}$ && $-\frac12
{\scriptstyle (D-1)(D-2)} y I[A^{\prime 2}] - \frac12
{\scriptstyle (D-1)^2 (D-2)} I^2[A^{\prime 2}]$ & \cr
\omit&height2pt&\omit&&\omit&\cr \tablerule
\omit&height2pt&\omit&&\omit&\cr && $P_{4b}$ && ${\scriptstyle
(D-1)^2 (D-2)} I[A' B']$ & \cr \omit&height2pt&\omit&&\omit&\cr
\tablerule \omit&height2pt&\omit&&\omit&\cr && $P_{7b}$ &&
${\scriptstyle (D-1)(D-2)} I[A' {\Delta C}']$ & \cr
\omit&height2pt&\omit&&\omit&\cr \tablerule
\omit&height2pt&\omit&&\omit&\cr && $P_{7c}$ && $-\frac12
{\scriptstyle (D-1)(D-2)} y I[A' {\Delta C}'] - \frac12
{\scriptstyle (D-1)^2 (D-2)} I^2[A' {\Delta C}']$ & \cr
\omit&height2pt&\omit&&\omit&\cr \tablerule
\omit&height2pt&\omit&&\omit&\cr \tablerule
\omit&height2pt&\omit&&\omit&\cr && \omit &&
$\!\!\!\!\!{\scriptstyle D (D-1) (D-2)} I[A' B'] - {\scriptstyle
(D-1) (D-2)} I[A' {\Delta B}'] -\frac12 {\scriptstyle (D-1) (D-2)}
y I[A^{\prime 2}]\!\!\!\!\!$ &\cr \omit&height2pt&\omit&&\omit&\cr
&& Total && $- \frac12 {\scriptstyle (D-1)^2 (D-2)} I^2[A^{\prime
2}] + {\scriptstyle (D-1) (D-2)} I[A' {\Delta C}']$ & \cr
\omit&height2pt&\omit&&\omit&\cr && \omit && $-\frac12
{\scriptstyle (D-1) (D-2)} y I[A' {\Delta C}'] - \frac12
{\scriptstyle (D-1)^2 (D-2)} I^2[A' {\Delta C}']$ & \cr
\omit&height2pt&\omit&&\omit&\cr \tablerule}}

\caption{Contributions acted upon by $\gamma_1 = (a a')^D H^4$.}

\label{gamma1}

\end{table}

\begin{table}

\vbox{\tabskip=0pt \offinterlineskip
\def\tablerule{\noalign{\hrule}}
\halign to390pt {\strut#& \vrule#\tabskip=1em plus2em&
\hfil#\hfil& \vrule#& \hfil#\hfil& \vrule#\tabskip=0pt\cr
\tablerule \omit&height4pt&\omit&&\omit&\cr &&$\!\!\!\!{\rm
Part}\!\!\!\!$ && $\!\!\!\! {\rm Contribution\ Acted\ upon\ by} \;
\gamma_2 \!\!\!\!$ & \cr \omit&height4pt&\omit&&\omit&\cr
\tablerule \omit&height2pt&\omit&&\omit&\cr && $P_{3b}$ && $y
A^{\prime 2}$ & \cr \omit&height2pt&\omit&&\omit&\cr \tablerule
\omit&height2pt&\omit&&\omit&\cr && $P_{3c}$ && $-\frac12 y^2
A^{\prime 2} -\frac{D}2 y I[A^{\prime 2}]$ & \cr
\omit&height2pt&\omit&&\omit&\cr \tablerule
\omit&height2pt&\omit&&\omit&\cr && $P_{4b}$ && ${\scriptstyle
(D-1)} y A' B'$ & \cr \omit&height2pt&\omit&&\omit&\cr \tablerule
\omit&height2pt&\omit&&\omit&\cr && $P_{7b}$ && $y A' {\Delta C}'$
& \cr \omit&height2pt&\omit&&\omit&\cr \tablerule
\omit&height2pt&\omit&&\omit&\cr && $P_{7c}$ && $-\frac12 y^2 A'
{\Delta C}' - \frac{D}2 y I[A' {\Delta C}']$ & \cr
\omit&height2pt&\omit&&\omit&\cr \tablerule
\omit&height2pt&\omit&&\omit&\cr \tablerule
\omit&height2pt&\omit&&\omit&\cr && Total && ${\scriptstyle D} y
A' B' - y A' {\Delta B}' - \frac12 y^2 A^{\prime 2} - \frac{D}2 y
I[A^{\prime 2}]$ & \cr \omit&height2pt&\omit&&\omit&\cr && \omit
&& $+ y A' {\Delta C}' - \frac{D}2 y I[A' {\Delta C}'] - \frac12
y^2 A' {\Delta C}'$ & \cr \omit&height2pt&\omit&&\omit&\cr
\tablerule}}

\caption{Contributions acted upon by $\gamma_2 = (a a')^{D-1} (a^2
+ a^{\prime 2}) H^4$.}

\label{gamma2}

\end{table}

\begin{table}

\vbox{\tabskip=0pt \offinterlineskip
\def\tablerule{\noalign{\hrule}}
\halign to390pt {\strut#& \vrule#\tabskip=1em plus2em&
\hfil#\hfil& \vrule#& \hfil#\hfil& \vrule#\tabskip=0pt\cr
\tablerule \omit&height4pt&\omit&&\omit&\cr &&$\!\!\!\!{\rm
Part}\!\!\!\!$ && $\!\!\!\! {\rm Contribution\ Acted\ upon\ by} \;
\gamma_3 \!\!\!\!$ & \cr \omit&height4pt&\omit&&\omit&\cr
\tablerule \omit&height2pt&\omit&&\omit&\cr && $P_{3b}$ &&
${\scriptstyle (D-1)} y A^{\prime 2} + y^2 (A^{\prime 2})'$ & \cr
\omit&height2pt&\omit&&\omit&\cr \tablerule
\omit&height2pt&\omit&&\omit&\cr && $P_{3c}$ && $- {\scriptstyle
D} y^2 A^{\prime 2} - \frac12 y^3 (A^{\prime 2})' -\frac12
{\scriptstyle (D-1) D} y I[A^{\prime 2}]$ & \cr
\omit&height2pt&\omit&&\omit&\cr \tablerule
\omit&height2pt&\omit&&\omit&\cr && $P_{4b}$ && ${\scriptstyle
(D-1)^2} y A' B' + {\scriptstyle (D-1)} y^2 (A' B')'$ & \cr
\omit&height2pt&\omit&&\omit&\cr \tablerule
\omit&height2pt&\omit&&\omit&\cr && $P_{7b}$ && ${\scriptstyle
(D-1)} y A' {\Delta C}' + y^2 (A' {\Delta C}')'$ & \cr
\omit&height2pt&\omit&&\omit&\cr \tablerule
\omit&height2pt&\omit&&\omit&\cr && $P_{7c}$ && $-{\scriptstyle D}
y^2 A' {\Delta C}' - \frac12 y^3 (A' {\Delta C}')' -\frac12
{\scriptstyle (D-1) D} y I[A' {\Delta C}']$ & \cr
\omit&height2pt&\omit&&\omit&\cr \tablerule
\omit&height2pt&\omit&&\omit&\cr \tablerule
\omit&height2pt&\omit&&\omit&\cr && \omit && ${\scriptstyle (D-1)
D} y A' B' + {\scriptstyle D} y^2 (A' B')' - {\scriptstyle (D-1)}
y A' {\Delta B}' - y^2 (A' {\Delta B}')'$ & \cr
\omit&height2pt&\omit&&\omit&\cr && Total && $- {\scriptstyle D}
y^2 A^{\prime 2} - \frac12 y^3 (A^{\prime 2})' -\frac12
{\scriptstyle (D-1) D} y I[A^{\prime 2}] + {\scriptstyle (D-1)} y
A' {\Delta C}'$ & \cr \omit&height2pt&\omit&&\omit&\cr && \omit &&
$\!\!\!\!+ y^2 (A' {\Delta C}')' - {\scriptstyle D} y^2 A' {\Delta
C}' - \frac12 y^3 (A' {\Delta C}')' - \frac12 {\scriptstyle (D-1)
D} y I[A' {\Delta C}']\!\!\!\!$ & \cr
\omit&height2pt&\omit&&\omit&\cr \tablerule}}

\caption{Contributions acted upon by $\gamma_3 = (a a')^{D-1} (a +
a')^2 H^4$.}

\label{gamma3}

\end{table}

\begin{table}

\vbox{\tabskip=0pt \offinterlineskip
\def\tablerule{\noalign{\hrule}}
\halign to390pt {\strut#& \vrule#\tabskip=1em plus2em&
\hfil#\hfil& \vrule#& \hfil#\hfil& \vrule#\tabskip=0pt\cr
\tablerule \omit&height4pt&\omit&&\omit&\cr &&$\!\!\!\!{\rm
Part}\!\!\!\!$ && $\!\!\!\! {\rm Contribution\ Acted\ upon\ by} \;
\delta \!\!\!\!$ & \cr \omit&height4pt&\omit&&\omit&\cr \tablerule
\omit&height2pt&\omit&&\omit&\cr && $P_{2+5}$ && $I^2[{\Delta B}
A'']$ & \cr \omit&height2pt&\omit&&\omit&\cr \tablerule
\omit&height2pt&\omit&&\omit&\cr && $P_{3a}$ && $-I^2[A A'']$ &
\cr \omit&height2pt&\omit&&\omit&\cr \tablerule
\omit&height2pt&\omit&&\omit&\cr && $P_{3d}$ && $\frac14
I^3[A^{\prime 2}]$ & \cr \omit&height2pt&\omit&&\omit&\cr
\tablerule \omit&height2pt&\omit&&\omit&\cr && $P_{4a}$ &&
$\frac12 I^2[B A'']$ & \cr \omit&height2pt&\omit&&\omit&\cr
\tablerule \omit&height2pt&\omit&&\omit&\cr && $P_{6a}$ &&
$\frac12 I^2[B A'']$ & \cr \omit&height2pt&\omit&&\omit&\cr
\tablerule \omit&height2pt&\omit&&\omit&\cr && $P_{7a}$ &&
$-I^2[{\Delta C} A'']$ & \cr \omit&height2pt&\omit&&\omit&\cr
\tablerule \omit&height2pt&\omit&&\omit&\cr && $P_{7d}$ &&
$\frac14 I^3[A' {\Delta C}']$ & \cr
\omit&height2pt&\omit&&\omit&\cr \tablerule
\omit&height2pt&\omit&&\omit&\cr \tablerule
\omit&height2pt&\omit&&\omit&\cr && Total && $2 I^2[{\Delta B}
A''] + \frac14 I^3[A^{\prime 2}] - I^2[{\Delta C} A''] + \frac14
I^3[A' {\Delta C}']$ & \cr \omit&height2pt&\omit&&\omit&\cr
\tablerule}}

\caption{Contributions acted upon by $\delta = (a a')^{D-2} (a^2 +
a^{\prime 2}) \nabla^2 \square$.}

\label{delta}

\end{table}

\begin{table}

\vbox{\tabskip=0pt \offinterlineskip
\def\tablerule{\noalign{\hrule}}
\halign to390pt {\strut#& \vrule#\tabskip=1em plus2em&
\hfil#\hfil& \vrule#& \hfil#\hfil& \vrule#\tabskip=0pt\cr
\tablerule \omit&height4pt&\omit&&\omit&\cr &&$\!\!\!\!{\rm
Part}\!\!\!\!$ && $\!\!\!\! {\rm Contribution\ Acted\ upon\ by} \;
\epsilon_1 \!\!\!\!$ & \cr \omit&height4pt&\omit&&\omit&\cr
\tablerule \omit&height2pt&\omit&&\omit&\cr && $P_{1b}$ && $2
{\scriptstyle (D-1)} I[A^{\prime 2}]$ & \cr
\omit&height2pt&\omit&&\omit&\cr \tablerule
\omit&height2pt&\omit&&\omit&\cr && $P_{2+5}$ && ${\scriptstyle
-(D-1)} I^2[A' {\Delta B}']$ & \cr
\omit&height2pt&\omit&&\omit&\cr \tablerule
\omit&height2pt&\omit&&\omit&\cr && $P_{3b}$ && $-2 I[A^{\prime
2}] + \frac{D}2 I^2[A^{\prime 2}]$ & \cr
\omit&height2pt&\omit&&\omit&\cr \tablerule
\omit&height2pt&\omit&&\omit&\cr && $P_{3c}$ && $y I[A^{\prime 2}]
+ {\scriptstyle (D-1)} I^2[A^{\prime 2}] - \frac{D}4 y
I^2[A^{\prime 2}] - \frac14 {\scriptstyle D (D-2)} I^3[A^{\prime
2}]$ & \cr \omit&height2pt&\omit&&\omit&\cr \tablerule
\omit&height2pt&\omit&&\omit&\cr && $P_{3d}$ && $\frac14
{\scriptstyle (D-2) (D-3)} I^3[A^{\prime 2}]$ & \cr
\omit&height2pt&\omit&&\omit&\cr \tablerule
\omit&height2pt&\omit&&\omit&\cr && $P_{4a}$ && $\frac12
{\scriptstyle (D-2) (D-3)} I^2[B A'']$ & \cr
\omit&height2pt&\omit&&\omit&\cr \tablerule
\omit&height2pt&\omit&&\omit&\cr && $P_{4b}$ && $-2 {\scriptstyle
(D-1)} I[A' B'] + \frac12 {\scriptstyle (D-1) D} I^2[A' B']$ & \cr
\omit&height2pt&\omit&&\omit&\cr \tablerule
\omit&height2pt&\omit&&\omit&\cr && $P_{6a}$ && $\frac12
{\scriptstyle (D-1) (D-2)} I^2[B A'']$ & \cr
\omit&height2pt&\omit&&\omit&\cr \tablerule
\omit&height2pt&\omit&&\omit&\cr && $P_{6b}$ && $2 I[A' B']$ & \cr
\omit&height2pt&\omit&&\omit&\cr \tablerule
\omit&height2pt&\omit&&\omit&\cr && $P_{6c}$ && $-y I[A' B'] -
{\scriptstyle (D-1)} I^2[A' B']$ & \cr
\omit&height2pt&\omit&&\omit&\cr \tablerule
\omit&height2pt&\omit&&\omit&\cr && $P_{7b}$ && $-2 I[A' {\Delta
C}'] + \frac{D}2 I^2[A' {\Delta C}']$ & \cr
\omit&height2pt&\omit&&\omit&\cr \tablerule
\omit&height2pt&\omit&&\omit&\cr && $P_{7c}$ && $\!\!\!\!\! y I[A'
{\Delta C}'] + {\scriptstyle (D-1)} I^2[A' {\Delta C}'] -
\frac{D}4 y I^2[A' {\Delta C}'] - \frac14 {\scriptstyle D (D-2)}
I^3[A' {\Delta C}'] \!\!\!\!\!$ & \cr
\omit&height2pt&\omit&&\omit&\cr \tablerule
\omit&height2pt&\omit&&\omit&\cr && $P_{7d}$ && $\frac14
{\scriptstyle (D-2) (D-3)} I^3[A' {\Delta C}']$ & \cr
\omit&height2pt&\omit&&\omit&\cr \tablerule
\omit&height2pt&\omit&&\omit&\cr \tablerule
\omit&height2pt&\omit&&\omit&\cr && \omit && $\!\!\!\!\!-2
{\scriptstyle (D-2)} I[A' {\Delta B}'] \!-\! y I[A' {\Delta B}']
\!-\! (\frac{5D-4}2) I^2[A' {\Delta B}'] \!+\! \frac{D^2}2 I^2[A'
B']\!\!\!\!\!$ & \cr \omit&height2pt&\omit&&\omit&\cr &&
$\!\!\!\!\! {\rm Total} \!\!\!\!\!$ && $- \frac{D}4 y
I^2[A^{\prime 2}] + {\scriptstyle (D-2)^2} I^2[A'' B] - \frac34
{\scriptstyle (D-2)} I^3[A^{\prime 2}] -2 I[A' {\Delta C}']$ & \cr
\omit&height2pt&\omit&&\omit&\cr && \omit && $\!\!\!\!\!+ y I[A'
{\Delta C}'] \!+\! (\frac{3D-2}2) I^2[A' {\Delta C}'] \!-\!
\frac{D}4 y I^2[A' {\Delta C}'] \!-\! \frac34 {\scriptstyle (D-2)}
I^3[A' {\Delta C}'] \!\!\!\!\!$ & \cr
\omit&height2pt&\omit&&\omit&\cr \tablerule}}

\caption{Contributions acted upon by $\epsilon_1 = (a a')^{D-1}
H^2 \nabla^2$.}

\label{epsilon1}

\end{table}

\begin{table}

\vbox{\tabskip=0pt \offinterlineskip
\def\tablerule{\noalign{\hrule}}
\halign to390pt {\strut#& \vrule#\tabskip=1em plus2em&
\hfil#\hfil& \vrule#& \hfil#\hfil& \vrule#\tabskip=0pt\cr
\tablerule \omit&height4pt&\omit&&\omit&\cr &&$\!\!\!\!{\rm
Part}\!\!\!\!$ && $\!\!\!\! {\rm Contribution\ Acted\ upon\ by} \;
\epsilon_2 \!\!\!\!$ & \cr \omit&height4pt&\omit&&\omit&\cr
\tablerule \omit&height2pt&\omit&&\omit&\cr && $P_{2+5}$ &&
$-{\scriptstyle (D-1)} I^2[A' {\Delta B}'] + y I[B A''] + I^2[A'
B']$ & \cr \omit&height2pt&\omit&&\omit&\cr \tablerule
\omit&height2pt&\omit&&\omit&\cr && $P_{3d}$ && $\frac14 y
I^2[A^{\prime 2}]$ & \cr \omit&height2pt&\omit&&\omit&\cr
\tablerule \omit&height2pt&\omit&&\omit&\cr && $P_{4a}$ &&
$\frac12 y I[B A'']$ & \cr \omit&height2pt&\omit&&\omit&\cr
\tablerule \omit&height2pt&\omit&&\omit&\cr && $P_{6a}$ &&
$\frac12 y I[B A'']$ & \cr \omit&height2pt&\omit&&\omit&\cr
\tablerule \omit&height2pt&\omit&&\omit&\cr && $P_{7d}$ &&
$\frac14 y I^2[A' {\Delta C}']$ & \cr
\omit&height2pt&\omit&&\omit&\cr \tablerule
\omit&height2pt&\omit&&\omit&\cr \tablerule
\omit&height2pt&\omit&&\omit&\cr && $\!\!\!\!\! {\rm Total}
\!\!\!\!\!$ && $\!\!\!\!\! 2 y I[B A''] + I^2[A' B'] \!\!\!\!\!$ &
\cr && \omit && $-{\scriptstyle (D-1)} I^2[A' {\Delta B}'] +
\frac14 y I^2[A^{\prime 2}] + \frac14 y I^2[A' {\Delta
C}']\!\!\!\!\!$ & \cr \omit&height2pt&\omit&&\omit&\cr
\tablerule}}

\caption{Contributions acted upon by $\epsilon_2 = (a a')^{D-2}
(a^2 + a^{\prime 2}) H^2 \nabla^2$.}

\label{epsilon2}

\end{table}

\begin{table}

\vbox{\tabskip=0pt \offinterlineskip
\def\tablerule{\noalign{\hrule}}
\halign to390pt {\strut#& \vrule#\tabskip=1em plus2em&
\hfil#\hfil& \vrule#& \hfil#\hfil& \vrule#\tabskip=0pt\cr
\tablerule \omit&height4pt&\omit&&\omit&\cr &&$\!\!\!\!{\rm
Part}\!\!\!\!$ && $\!\!\!\! {\rm Contribution\ Acted\ upon\ by} \;
\epsilon_3 \!\!\!\!$ & \cr \omit&height4pt&\omit&&\omit&\cr
\tablerule \omit&height2pt&\omit&&\omit&\cr && $P_{2+5}$ && $y^2
A'' B + {\scriptstyle (D-1)} y I[A'' B] + {\scriptstyle (D-2)}
I^2[A' B'] + y I[A^{\prime 2}]$ & \cr
\omit&height2pt&\omit&&\omit&\cr \tablerule
\omit&height2pt&\omit&&\omit&\cr && $P_{3d}$ && $(\frac{D-1}4) y
I^2[A^{\prime 2}] + \frac14 y^2 I[A^{\prime 2}]$ & \cr
\omit&height2pt&\omit&&\omit&\cr \tablerule
\omit&height2pt&\omit&&\omit&\cr && $P_{4a}$ && $(\frac{D-1}2) y
I[A'' B] + \frac12 y^2 A'' B$ & \cr
\omit&height2pt&\omit&&\omit&\cr \tablerule
\omit&height2pt&\omit&&\omit&\cr && $P_{6a}$ && $(\frac{D-1}2) y
I[A'' B] + \frac12 y^2 A'' B$ & \cr
\omit&height2pt&\omit&&\omit&\cr \tablerule
\omit&height2pt&\omit&&\omit&\cr && $P_{7d}$ && $(\frac{D-1}4) y
I^2[A' {\Delta C}'] + \frac14 y^2 I[A' {\Delta C}']$ & \cr
\omit&height2pt&\omit&&\omit&\cr \tablerule
\omit&height2pt&\omit&&\omit&\cr \tablerule
\omit&height2pt&\omit&&\omit&\cr && Total && $2 y^2 A'' B + 2
{\scriptstyle (D-1)} y I[A'' B] + {\scriptstyle (D-2)} I^2[A' B']
+ y I[A^{\prime 2}]$ & \cr \omit&height2pt&\omit&&\omit&\cr &&
\omit && $\!\!\!\!+ (\frac{D-1}4) y I^2[A^{\prime 2}] + \frac14
y^2 I[A^{\prime 2}] + (\frac{D-1}4) y I^2[A' {\Delta C}'] +
\frac14 y^2 I[A' {\Delta C}']\!\!\!\!$ & \cr
\omit&height2pt&\omit&&\omit&\cr \tablerule}}

\caption{Contributions acted upon by $\epsilon_3 = (a a')^{D-2} (a
+ a')^2 H^2 \nabla^2$.}

\label{epsilon3}

\end{table}

\begin{table}

\vbox{\tabskip=0pt \offinterlineskip
\def\tablerule{\noalign{\hrule}}
\halign to390pt {\strut#& \vrule#\tabskip=1em plus2em&
\hfil#\hfil& \vrule#& \hfil#\hfil& \vrule#\tabskip=0pt\cr
\tablerule \omit&height4pt&\omit&&\omit&\cr &&$\!\!\!\!{\rm
Part}\!\!\!\!$ && $\!\!\!\! {\rm Contribution\ Acted\ upon\ by} \;
\zeta \!\!\!\!$ & \cr \omit&height4pt&\omit&&\omit&\cr \tablerule
\omit&height2pt&\omit&&\omit&\cr && $P_{1a}$ && $I^2[A A'']$ & \cr
\omit&height2pt&\omit&&\omit&\cr \tablerule
\omit&height2pt&\omit&&\omit&\cr && $P_{2+5}$ && $-2 I^2[{\Delta
B} A''] + \frac12 I^3[A' {\Delta B}']$ & \cr
\omit&height2pt&\omit&&\omit&\cr \tablerule
\omit&height2pt&\omit&&\omit&\cr && $P_{3a}$ && $I^2[A A'']$ & \cr
\omit&height2pt&\omit&&\omit&\cr \tablerule
\omit&height2pt&\omit&&\omit&\cr && $P_{3d}$ && $-\frac12
I^3[A^{\prime 2}] + (\frac{D-2}8) I^4[A^{\prime 2}]$ & \cr
\omit&height2pt&\omit&&\omit&\cr \tablerule
\omit&height2pt&\omit&&\omit&\cr && $P_{4a}$ && $- I^2[A'' B] +
(\frac{D-2}4) I^3[A'' B]$ & \cr \omit&height2pt&\omit&&\omit&\cr
\tablerule \omit&height2pt&\omit&&\omit&\cr && $P_{6a}$ && $-
I^2[A'' B] - (\frac{D-2}4) I^3[A'' B]$ & \cr
\omit&height2pt&\omit&&\omit&\cr \tablerule
\omit&height2pt&\omit&&\omit&\cr && $P_{6d}$ && $\frac12 I^3[A'
B']$ & \cr \omit&height2pt&\omit&&\omit&\cr \tablerule
\omit&height2pt&\omit&&\omit&\cr && $P_{7a}$ && $I^2[{\Delta C}
A'']$ & \cr \omit&height2pt&\omit&&\omit&\cr \tablerule
\omit&height2pt&\omit&&\omit&\cr && $P_{7d}$ && $-\frac12 I^3[A'
{\Delta C}'] + (\frac{D-2}8) I^4[A' {\Delta C}']$ & \cr
\omit&height2pt&\omit&&\omit&\cr \tablerule
\omit&height2pt&\omit&&\omit&\cr \tablerule
\omit&height2pt&\omit&&\omit&\cr && Total && $-4 I^2[A'' {\Delta
B}] + I^3[A' {\Delta B}'] + (\frac{D-2}8) {\scriptstyle }
I^4[A^{\prime 2}]$ & \cr \omit&height2pt&\omit&&\omit&\cr && \omit
&& $+ I^2[{\Delta C} A''] -\frac12 I^3[A' {\Delta C}'] +
(\frac{D-2}8) {\scriptstyle } I^4[A' {\Delta C}']$ & \cr
\omit&height2pt&\omit&&\omit&\cr \tablerule}}

\caption{Contributions acted upon by $\zeta = (a a')^{D-2}
\nabla^4$.}

\label{zeta}

\end{table}

\end{document}